\documentclass[]{jfm}

\usepackage{graphicx}
\usepackage{newtxtext}
\usepackage{newtxmath}
\usepackage{natbib}
\usepackage{hyperref}
\hypersetup{
    colorlinks = true,
    urlcolor   = blue,
    citecolor  = black,
    linkcolor=blue
}

\newcommand{\RomanNumeralCaps}[1]

\newcommand\citecol[1]{\hypersetup{citecolor=black}#1\hypersetup{citecolor=black}}

\defcitealias{kafiabad_savva_vanneste_2019}{KSV}
\defcitealias{cox_kafiabad_vanneste_2023}{CKV}

\usepackage{xcolor, soul}

\usepackage[normalem]{ulem}

\def\bk{\boldsymbol{k}}
\def\bK{\boldsymbol{K}}
\def\bx{\boldsymbol{x}}
\def\by{\boldsymbol{y}}
\def\bX{\boldsymbol{X}}
\def\bU{\boldsymbol{U}}
\def\bu{\boldsymbol{u}}
\def\bc{\boldsymbol{c}}
\newcommand{\pa}[1]{\partial_{#1}}
\def\gx{\bnabla_{\bx}}
\def\gy{\bnabla_{\by}}
\def\gk{\bnabla_{\bk}}
\def\omi{\omega_{0}}
\def\omf{\omega_{1}}
\def\be{\boldsymbol{e}}
\def\DH{\Delta H}
\def\iOmega{\mathit{\Omega}}
\def\Rsw{R_{\text{sw}}}
\def\RB{R_{\text{B}}}
\def\RBpk{R_{\text{B},kk}'}
\def\RBpp{R_{\text{B},\phi \phi}'}
\newcommand{\num}[1]{^{(#1)}}
\def\d{\text{d}}
\def\D{\mathsfi{D}}
\def\bD{\mathsfbi{D}}

\def\iTheta{\mathit{\Theta}}
\def\omin{\omega_{\text{in}}}

\def\T{\mathrm{T}}
\def\cc{\mathrm{c.c.}}

\def\gX{\bnabla_{\bX}}
\def\bphi{\boldsymbol{\phi}}
\def\mmu{\boldsymbol{\mathsfbi{\mu}}}
\def\mB{\mathsfbi{B}}
\newcommand{\Dv}[1]{\mathcal{D}_{#1}}
\newcommand{\mcirc}[1]{\text{\textcircled{#1}}}
\def\bcin{\bc_{\text{in}}}
\def\balpha{\boldsymbol{\ell}}
\def\Nb{\bar{N}}

\def\i{\mathrm{i}}
\def\e{\mathrm{e}}

\title{Inhomogeneity-induced wavenumber diffusion}

\author{Michael R. Cox\aff{1}
  \corresp{\email{michael.cox@ed.ac.uk}},
  Hossein A. Kafiabad\aff{2}
 \and Jacques Vanneste\aff{1}}

\affiliation{\aff{1}School of Mathematics and Maxwell Institute for Mathematical Sciences, University of Edinburgh,
Edinburgh EH9 3FD, UK \\
\aff{2}Department of Mathematical Sciences, 
Durham University, 
Durham DH1 3LE, UK }

\begin{document}
\maketitle

\begin{abstract}
Inertia-gravity waves are scattered by background flows as a result of Doppler shift by a non-uniform velocity. In the WKB regime, the scattering process reduces to a diffusion in spectral space. Other inhomogeneities the waves encounter, such as density variations, also cause scattering and spectral diffusion. We generalise the spectral diffusion equation to account for these inhomogeneities. We apply the result to the rotating shallow water system, for which height inhomogeneities arise from velocity inhomogeneities through geostrophy, and to the Boussinesq system for which buoyancy inhomogeneities arise similarly. We compare the contributions that height and buoyancy variations make to the spectral diffusion with the contribution of the Doppler shift.  In both systems, we find regimes where all contributions are significant. We support our findings with exact solutions of the diffusion equation and  with ray tracing simulations in the shallow water case.
\end{abstract}





\section{Introduction}


Inertia-gravity waves (IGWs) propagate  in the atmosphere and ocean under the restoring forces of buoyancy and  Coriolis effect. As they propagate, they encounter and interact with a variety of inhomogeneities, including background flows, topography and other waves. These inhomogeneities refract, reflect and, in the case of background flows, advect the waves. For example, internal tides -- IGWs generated at the semi-diurnal and diurnal frequencies by astronomical forcing -- are advected and refracted by background flows
(e.g.\ \citeauthor*{park2006internal} \citeyear{park2006internal}; \citeauthor*{rainville2006propagation} \citeyear{rainville2006propagation}; \citeauthor{chavanne2010surface} \citeyear{chavanne2010surface}; \citeauthor*{Pan_Haley_Lermusiaux_2021} \citeyear{Pan_Haley_Lermusiaux_2021})
and reflected by bottom topography
(e.g.\ \citeauthor{muller1992scattering} \citeyear{muller1992scattering}; \citecol{\citeauthor{BÜHLER_HOLMES-CERFON_2011} \citeyear{BÜHLER_HOLMES-CERFON_2011}}; \citeauthor{kelly2013geography} \citeyear{kelly2013geography}; \citeauthor*{lahaye2020internal} \citeyear{lahaye2020internal}; \citeauthor{Pan_Haley_Lermusiaux_2021} \citeyear{Pan_Haley_Lermusiaux_2021}).
IGWs are also affected by each other \citep[e.g.][]{eden2014energy}. 
In the limit of weak, repeated interactions, wave energy is redistributed in spectral space in a scattering process which can be described statistically.

Kinetic equations derived from the governing fluid equations provide this statistical description of scattering by weak, slowly evolving, random inhomogeneities (see \citet*{Hasselmann66} and \citet*{RYZHIK1996327} for general formulations). For example, \citet{Hasselmann66} and \citet{ARDHUIN_HERBERS_2002} derive a kinetic equation for surface waves scattered by gently-sloping bottom topography; \citet*{Eden2019} derive a scattering equation for IGWs interacting with a weak, random, slowly evolving wave field \citep*[see also][]{Eden2020}; and \citet*{dani-v16}, \citet*{savv-vann18} and \citet*{savva_kafiabad_vanneste_2021} derive kinetic equations for near-inertial waves, internal tides and IGWs respectively,  scattered by weak, slowly evolving turbulence. 
A key feature of scattering is that if the scattering inhomogeneities evolve sufficiently slowly compared with the waves, wave frequency is preserved.

This paper concerns the WKB limit of large-scale inhomogeneities scattering small-scale waves, in which scattering reduces to diffusion in spectral space. The corresponding induced-diffusion equation can be derived by invoking conservation of wave action density, which holds in the WKB limit. This was first done by \citet*{mccomas_bretherton_1977} in the context of wave-wave interactions using the ray equations -- the characteristic equations for action conservation. (\citet{mccomas_bretherton_1977} note that their derivation also applies to diffusion induced by low-frequency currents. A flow-induced diffusivity is first discussed in \citet*{MullerOlbers} and expanded on in \citeauthor{Mueller_1976} (\citeyear{Mueller_1976}, \citeyear{MULLER197749}).)
Recently, alternative derivations for the diffusion equation start with the conservation of wave action and use multi-scale asymptotics. They have been carried out for weak geostrophic flows scattering 1) IGWs in a 3D Boussinesq system (\citeauthor*{kafiabad_savva_vanneste_2019} \citeyear{kafiabad_savva_vanneste_2019}, hereafter \citetalias{kafiabad_savva_vanneste_2019}); 2) Poincaré waves in a rotating shallow water system \citep*{dong_buhler_smith_2020} and; 3) 
deep-water surface waves \citep{villas_boas_young_2020}. In all these derivations, the geostrophic flow impacts wave propagation solely through the Doppler shift term of the wave dispersion relation. \citet{mccomas_bretherton_1977} and \citet{Savva2020} show that induced-diffusion is the WKB limit of a scattering integral for wave-wave and wave-flow interactions respectively. \citet{barkan2023oceanic} investigate the relevance of diffusion theories in a realistic ocean simulation.


The aims of this paper are twofold: 1) we argue at the level of the dispersion relation  that inhomogeneities other than Doppler shift can be significant in scattering waves \textcolor{black}{in the WKB regime} (\S\ref{sec:scalingarguments}); and 2) we derive the spectral \textcolor{black}{diffusivity} induced by any weak inhomogeneity (\S\ref{sec:GD}), thus generalising the original result of \citet{mccomas_bretherton_1977}. \textcolor{black}{For, say, waves scattered by bottom topography \citecol{\citep[e.g.][]{Hasselmann66,muller1992scattering, ARDHUIN_HERBERS_2002}}, it is widely appreciated that scattering mechanisms other than Doppler shift are significant. However, this point has at times been overlooked in the study of wave scattering by mean flows as we detail below.}



In \S\ref{sec:scalingarguments} we use scaling arguments to compare the Doppler shift term of the dispersion relation to two ofttimes neglected inhomogeneities: height fluctuations in a rotating shallow water system, and  buoyancy gradients in a Boussinesq system. The height fluctuation effects are neglected in \citet{dong_buhler_smith_2020} and the buoyancy fluctuation term is neglected in previous work of the authors, \citetalias{kafiabad_savva_vanneste_2019} and \citet*[][hereafter \citetalias{cox_kafiabad_vanneste_2023}]{cox_kafiabad_vanneste_2023}. For both systems, we find regimes where these inhomogeneities can be significant. Doppler shift and vertical flow buoyancy gradients are accounted for in tidal ray tracing implemented by \cite{chavanne2010surface}. They find both to be important and, in their simulations, refraction by these gradients is more significant in the transfer of wave energy than advection by Doppler shift. (The ray tracing formulation used by \citet{chavanne2010surface} is outlined in \citet{Olbers1981WKB} for IGWs propagating in geostrophic flows with vertical and horizontal shears.) Doppler shift, refraction through a background flow, buoyancy gradients and topography are all found to be significant in a coupled set of tidal equations derived by \citet{Pan_Haley_Lermusiaux_2021}.

\textcolor{black}{Our analysis throughout this paper applies to weak scattering of waves by inhomogeneities. In practice, this means that the scattering inhomogeneities induce small perturbations to the wave frequencies. The wave amplitudes are assumed small enough that linear wave theory applies and the wave feedback on the background flow is negligible.}


In \S\ref{sec:GD}, we impose further statistical assumptions on the inhomogeneities, assume they are slowly time dependent and follow the perturbation expansion of \citetalias{kafiabad_savva_vanneste_2019} to reach a general diffusion equation for any inhomogeneity. 
We evaluate the general diffusivity for our two example systems  \textcolor{black}{(the method is similar for both systems and so the Boussinesq evaluation is relegated to appendix \ref{app:boussdiff}). We then} revise the scaling arguments of the previous section. For the shallow water system, we support our analysis with ray tracing simulations (\S\ref{sec:rays}), finding good agreement with the exact solution for 2D wave action given in \citet{villas_boas_young_2020}. For the Boussinesq system, we find the forced equilibrium spectrum of wave energy. We also evaluate the Boussinesq diffusivity for a typical quasi-geostrophic flow. \textcolor{black}{We find the revised importance estimate of buoyancy fluctuation to Doppler shift effects obeys negative power laws in horizontal wavenumber and frequency as explained in appendix \ref{app:RBpowers}.}

\textcolor{black}{In \S\ref{sec:topography}, we evaluate the general diffusivity for scattering by topography of 1) rotating shallow water waves (\ref{app:RSWscatt}) and 2) surface waves propagating on a background current (\ref{app:SWscatt}). This is proof of concept that our general diffusivity has applications outside of waves scattered by mean flows.}

\textcolor{black}{Section \ref{sec:discussion} is a discussion of our results including possible applications and limitations. For completeness, we derive wave action conservation for a rotating shallow water system with significant height fluctuation effects in appendix \ref{app:wkb}. Lists of key symbols are included in appendix \ref{app:glossaries}.}


\section{Scaling arguments}\label{sec:scalingarguments}



In this section, we introduce a general inhomogeneity term into the wave dispersion relation for waves in the WKB regime. This term encompasses Doppler shift by a background flow but can include additional inhomogeneities. We motivate the inclusion of two such inhomogeneities: flow-induced height fluctuations in a rotating shallow water system and vertical buoyancy gradients in a 3D Boussinesq system. We find regimes where the additional inhomogeneities are significant, at points dominant. This has implications to ray tracing simulations which do not always take into account inhomogeneities other than Doppler shift and motivates the general diffusion equation derivation of \S\ref{sec:GD}, because previous diffusion equations of the form first introduced by \citet{mccomas_bretherton_1977} have only considered Doppler shift (by a background flow or long waves) as the scattering mechanism.

Our starting point is the conservation of wave action $a$ in $(\bx, \bk)$ space in the WKB limit of small-scale waves scattered by large-scale inhomogeneities:
\begin{equation}\label{actioncons}
    \pa{t} a + \gk \omega \bcdot \gx a - \gx \omega \bcdot \gk a = 0.
\end{equation}
\textcolor{black}{(This is derived for the rotating shallow water system in appendix \ref{app:wkb} but holds more generally.)} The absolute frequency of the waves $\omega$ is the sum of a homogeneous part, $\omi$, and a weak, inhomogeneity-induced part, $\omf$:
\begin{equation}\label{totfreq}
    \omega(\bx, \bk, t) = \omi(\bk) + \omf(\bx, \bk, t),
\end{equation}
with $\omf \ll \omi$. We call $\omi$ the \textit{bare} frequency and assume that it varies sufficiently slowly in time and space as to be considered constant. The frequency correction $\omf$ includes the Doppler shift $\bU \bcdot \bk$ \textcolor{black}{induced by a background velocity $\bU$}, and any other inhomogeneities. We consider two systems where other inhomogeneities inevitably arise in the presence of Doppler shift.

\subsection{Shallow water}\label{sec:scale_sw}

\begin{figure}
  \centerline{\includegraphics{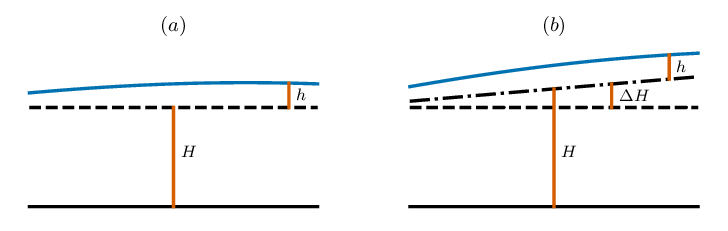}}
  \caption{Sketch of flat-bottom shallow water set-up. In both panels, the free surface is the solid blue line, the flat bottom is the solid black line and the dashed black line is the constant mean height $\bar{H}$. The wave perturbations are given by $h$. The height of the layer in the absence of waves is $H$ which is equal to $\bar{H}$ in panel (a) whilst in panel (b), it includes geostrophic height corrections $\DH$, given by the dashed-dotted line.}
\label{fig:SWpaper}
\end{figure}

We consider Poincar\'e waves propagating in rotating shallow water  with flat bottom and a background geostrophic flow. The background velocity $\bU = (U, V, 0)$ and height $H$ are related through the geostrophic balance
\begin{equation}\label{geobalSW}
    f \be_z \times \bU = - g \gx \DH,
\end{equation}
where $f$ is the Coriolis frequency, $g$ the gravitational constant and $\be_z$ the unit vertical vector. Here we have introduced $\DH$, the geostrophic perturbation to the mean height $\bar H$ such that $H = \bar H + \DH$ (see figure \ref{fig:SWpaper}). $\DH$ and hence $\bU$ is assumed to vary slowly in time and space compared with the period and wavelength of the Poincar\'e waves. For completeness we verify that the action conservation \eqref{actioncons} holds in this case in appendix \ref{app:wkb}.

Assuming that  $\DH \ll \bar{H}$, the frequency of waves with wavevector $\bk = (k_1, k_2)$ can be approximated as 
\begin{equation}\label{SWtot_2} 
    \omega = (f^2 + g (\bar{H} + \DH) k_h^2)^{1/2} + \bU \bcdot \bk \approx \omi + \frac{g \DH k_h^2}{2 \omi} + \bU \bcdot \bk,
\end{equation}
where $k_h = (k_1^2 + k_2^2)^{1/2}$ is the wavenumber and  $\omi = (f^2 + g \bar{H} k_h^2)^{1/2}$ is the intrinsic frequency for constant height  $\bar{H}$.
%
Thus, there are two contributions to the frequency inhomogeneity, 
\begin{equation}\label{omfSW}
    \omf = \underbrace{\frac{g \DH k_h^2}{2 \omi}}_{\text{height fluctuation}} + \underbrace{\vphantom{\frac{g \DH k_h^2}{2 \omi}} \bU \bcdot \bk}_{\text{Doppler shift}},
\end{equation}
and hence two contributions to the scattering.

\begin{figure}
  \centerline{\includegraphics{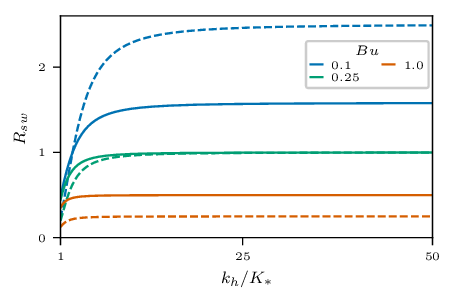}}
  \caption{Ratio $\Rsw$ (solid lines) defined by \eqref{ratioSW}, the estimated relative importance of height fluctuation and Doppler shift terms in the shallow water system,  plotted for different values of the Burger number $Bu$ against non-dimensionalised horizontal wavenumber $k_h/K_* > 1$. QG flow is $Bu = 1$. $Bu = 0.1, 0.25$ are associated with the PG regime. Dashed lines are the square of this ratio, $\Rsw'$ \eqref{Rswprime}, which gives a more accurate relative importance ratio for the diffusion regime, as shown in \S\ref{sec:GDSW}.
  }
\label{fig:ratio2}
\end{figure}

We now compare the relative size of the terms in \eqref{omfSW} to argue that height fluctuation effects can be just as important as Doppler shift in ray tracing and induced diffusion. We introduce the characteristic \textcolor{black}{background} flow speed $U_*$, characteristic horizontal wavenumber $K_*$ and the flow Burger number
\begin{equation}\label{burgereqn}
    Bu = \frac{g\bar{H}}{f^2} K_*^2.
\end{equation}
Using the geostrophic balance \eqref{geobalSW} and expressing $\omi$ in terms of $Bu$, the height fluctuation term scales like
\begin{equation}
    \frac{g \DH k_h^2}{2 \omi} \sim \frac{ U_* k_h^2 }{2 K_* (1 + Bu (k_h/K_*)^2))^{1/2}}.
\end{equation}
Then, the ratio between the height fluctuation and Doppler shift terms is given by
\begin{equation}\label{ratioSW}
    \Rsw = \frac{\text{height fluctuation}}{\text{Doppler shift}} \sim \frac{k_h/K_*}{2(1 + Bu \, (k_h/K_*)^2)^{1/2}}.
\end{equation}
In figure \ref{fig:ratio2}, we plot the ratio $\Rsw$ given by $\eqref{ratioSW}$ against non-dimensionalised wavenumber $k_h/K_*$ for different realistic values of the Burger number: $Bu = O(1)$ corresponds to the quasi-geostrophy (QG) regime and $Bu \ll 1$ to the planetary geostrophy (PG) regime.
We take the limit of \eqref{ratioSW} for large $k_h/K_*$, a necessary condition for \eqref{actioncons} to hold,
\begin{equation}
    \Rsw \rightarrow Bu^{-1/2}/2, \quad k_h/K_* \gg 1.
\end{equation}
We see that for $Bu = 0.25$, $\Rsw \rightarrow 1$ and the height fluctuation and Doppler shift terms in \eqref{omfSW} have the same magnitude.
For $Bu = 1$ corresponding to QG flow, $\Rsw \rightarrow 0.5$.
For smaller $Bu$ associated with PG, height fluctuations dominate over Doppler shift. For larger $Bu$, Doppler shift dominates over height fluctuations.


In \S\ref{sec:GDSW}, we show a better estimate for the relative importance of height fluctuation to Doppler shift effects in the diffusion regime is $\Rsw' = \Rsw^2$, which is also plotted in figure \ref{fig:ratio2}.


\subsection{3D Boussinesq}\label{sec:scalingarg_bouss}
\citet{chavanne2010surface} compare the effect of refraction through flow buoyancy gradients to that of Doppler shift on internal tide propagation. Our set-up is similar to theirs -- we consider 3D IGWs of wavevector $\bk = (k_1, k_2, k_3)$ propagating in a geostrophic flow $\bU = (U, V, 0)$ with buoyancy $B$. Unlike the tides, our waves are not in hydrostatic balance.

We first consider waves propagating with no background flow buoyancy gradients. The dispersion relation is
\begin{equation}\label{omtotBous}
    \omega = \left(f^2 \cos^2 \theta + N^2 \sin^2 \theta\right)^{1/2} + \bU \bcdot \bk.
\end{equation}
The intrinsic frequency is the first term, dependent on $\theta$, the angle between $\bk$ and the vertical, and $N$ is the buoyancy frequency. With uniform vertical buoyancy gradients, $N^2 = \Nb^2 = \mathrm{const}$, the bare frequency coincides with the intrinsic frequency. To obtain \eqref{omtotBous}, a WKB ansatz is substituted into the 3D Boussinesq equations. We omit the full derivation, but it follows the same method as appendix \ref{app:wkb} for the shallow water system. (See also \textcolor{black}{derivations of action conservation with vertical buoyancy gradients by 1) \citecol{\citet{Bretherton66}} for gravity waves in a shear flow without the effect of rotation and; 2) \citecol{\citet{Pan_Haley_Lermusiaux_2021}} for internal tides in hydrostatic balance}.)


We make a rough argument for the inclusion of inhomogeneous vertical buoyancy gradients associated with the geostrophic flow in \eqref{omtotBous}. Horizontal geostrophic balance and vertical hydrostatic balance lead to the thermal wind balance
\begin{equation}\label{thermalwind}
    \boldsymbol{f} \times \pa{z}\bU = \boldsymbol{\nabla}_{\bx,h} B.
\end{equation}
The $h$-subscript indicates a purely horizontal gradient. This means horizontal flow buoyancy gradients are induced by vertical shear. If the flow's vertical shear is nonlinear, vertical buoyancy gradients are also induced.

In deriving \eqref{omtotBous}, the buoyancy frequency $N^2$ only appears in the equation for wave buoyancy $b$ which is given, for negligible background flow buoyancy gradients, by
\begin{equation}\label{wavebuoy}
    \pa{t} b + \bU \bcdot \gx b + N^2 w = 0.
\end{equation}
Here, $w$ is the vertical wave velocity and $N^2 = \Nb^2$. Gradients in $\bU$ have been neglected under the WKB ansatz \citep[see, for example,][]{Olbers1981WKB}.

The full wave velocity is $\bu = (u, v, w)$. Introducing flow buoyancy gradients such that $B = B(\bx)$, \eqref{wavebuoy} becomes
\begin{equation}
    \pa{t} b + \bU \bcdot \gx b + \bu_h \bcdot \boldsymbol{\nabla}_{\bx,h} B + N^2 w = 0,
\end{equation}
where
\begin{equation}\label{N2}
    N^2 = \Nb^2 + \pa{z} B
\end{equation}
\textcolor{black}{and $\bu_h = (u,v,0)$ is the horizontal wave velocity.} This is (A5) of appendix a of  \citet{chavanne2010surface}. The vertical buoyancy gradient $\pa{z}B$ acts as a perturbation to $\Nb^2$. If horizontal gradients are neglected, then \eqref{omtotBous} remains the same with variable $N^2$ defined by \eqref{N2}.

Horizontal buoyancy gradients introduce $u$ and $v$ terms into the buoyancy equation. This changes the nature of the eigenvalue problem for $\omega$. It can be shown, as discussed in \citet{chavanne2010surface}, that the perturbations to the total frequency \eqref{omtotBous} caused by these $u$ and $v$ terms are \textcolor{black}{imaginary}.
This corresponds to growing and decaying wave modes which exchange energy with the background flow. This \textcolor{black}{shear} instability is because buoyancy is a non-conservative force in a baroclinic fluid \citep{Jones2001}. Following \textcolor{black}{\citecol{\citet{Bretherton66}}}, \citet{Olbers1981WKB} and \citet{chavanne2010surface}, we neglect these horizontal gradients \textcolor{black}{which amounts to an assumption of large Richardson number} and leave exploration of this instability and its interaction with wavevector diffusion to future work.

Accounting for weak flow-induced vertical buoyancy gradients,  we expand the dispersion relation \eqref{omtotBous} with $N^2$ given by \eqref{N2} to obtain
\begin{equation}\label{omtotBous2}
    \omega = \underbrace{\vphantom{\frac{\pa{z}B \sin^2 \theta}{2\omega_0}}\left(f^2 \cos^2 \theta + \Nb^2 \sin^2 \theta\right)^{1/2}}_{\omi} + \underbrace{\frac{\pa{z}B \sin^2 \theta}{2\omega_0} + \bU \bcdot \bk}_{\omf},
\end{equation}
valid for small $\pa{z}B/\Nb^2$. We compare the size of the Doppler shift and buoyancy gradient terms in $\omf$.
We introduce the background flow aspect ratio $\alpha$ defined by
\begin{equation}\label{aspect}
    \alpha = \frac{L_*}{H_*} = \frac{K_{v*}}{K_*},
\end{equation}
for characteristic length scales $L_* = 1/K_*$ and $H_* = 1/K_{v*}$ in the horizontal and vertical respectively.
Using the thermal wind relation \eqref{thermalwind} and aspect ratio, the buoyancy fluctuation term scales like
\begin{equation}\label{boussmag}
    \frac{\pa{z} B \sin^2 \theta}{2 \omi} \sim 
    \frac{\alpha^2 U_* K_* \sin^2 \theta}{2 (\cos^2 \theta + (\Nb^2/f^2)\sin^2 \theta)^{1/2}}.
\end{equation}
Thus, the ratio $\RB$ between the buoyancy fluctuation and Doppler shift terms is roughly
\begin{equation}\label{ratioB}
    \RB = \frac{\text{buoyancy fluctuation}}{\text{Doppler shift}} \sim \frac{\alpha^2 \sin \theta}{2(\cos^2 \theta+(\Nb^2/f^2) \sin^2 \theta)^{1/2}(k/K_*)},
\end{equation}
where $k = k_h/\sin\theta$ is the wavenumber magnitude. It is instructive to consider $\RB$ as a function of non-dimensionalised frequency and horizontal wavenumber, $\omega_0/f$ and $k_h/K_*$:
\begin{equation}\label{ratioB2}
    \RB \sim \frac{1}{2} \, \frac{\alpha^2}{\Nb^2/f^2 - 1} \, \frac{(\omi/f)^2 - 1}{\omi/f} \, \frac{1}{k_h/K_*} \approx \frac{1}{2} \, \left(\frac{\alpha}{\Nb/f}\right)^2 \, \frac{(\omi/f)^2 - 1}{\omi/f} \, \frac{1}{k_h/K_*}.
\end{equation}
The second expression holds for $(\bar{N}/f)^2 \gg 1$, which is true in both the atmosphere and ocean. At $\omega_0 = f$, $\RB = 0$. This is to be expected: buoyancy effects are absent for vertically-propagating inertial waves. The ratio attains a maximum value of $\RB \sim \alpha^2(2(\bar{N}/f)(k_h/K_*))^{-1}$ when $\omi = \bar{N}$. The ratio decays as $(k_h/K_*)^{-1}$. In the WKB regime we consider, $k_h/K_* \gg 1$ and so it appears justified to assume the Doppler shift term dominates. However, for large values of $\alpha$ or when considering the lower limit of the WKB regime, this may not be the case.

\begin{figure}
  \centerline{\includegraphics{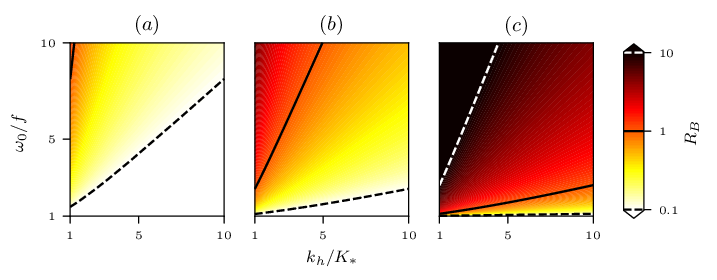}}
  \caption{Ratio $\RB$ as given in \eqref{ratioB2}, the relative importance of buoyancy fluctuation and Doppler shift terms in the full Boussinesq system, against non-dimensionalised frequency $\omega_0/f$ and horizontal wavenumber $k_h/K_*$. Panels $(a)$, $(b)$ and $(c)$ correspond to $\alpha = \bar{N}/(2f), \, \bar{N}/f,$ and $3 \bar{N}/f$ respectively, with $(\bar{N}/f)^2 \gg 1$. Contours are shown for $\RB = 0.1$ (dotted black), $1$ (solid black) and $10$ (dotted white).}
\label{fig:ratioBcontour}
\end{figure}

Figure \ref{fig:ratioBcontour} shows the ratio \eqref{ratioB2} for $(\bar{N}/f)^2 \gg 1$ and three values of $\alpha/(\bar{N}/f)$, \textcolor{black}{including} $\alpha \sim \bar{N}/f$, a realistic regime for geostrophic turbulence. Contours of $\RB = 0.1, 1$ and $10$, corresponding to a negligible, balanced and dominant buoyancy term respectively, are given for each value of $\alpha$. The buoyancy term can be equal to or greater than the Doppler shift term, namely for high aspect ratio $\alpha$, higher frequencies and lower wavenumbers.

\section{Diffusion regime}\label{sec:GD}
In this section, we impose statistical assumptions on slowly evolving, weak inhomogeneities to derive a spectral diffusion equation from the action conservation equation \eqref{actioncons}. We expand the derivation of \citetalias{kafiabad_savva_vanneste_2019} to include any total frequency of the form \eqref{totfreq} assuming 1) the inhomogeneity is weak enough, that is,  $\omf \ll \omi$ (this is the weak interaction limit of statistical scattering theories such as \citet{Hasselmann66}); 2) the bare frequency varies slowly over $\bx$; and 3) the inhomogeneity can be well modelled by a statistically homogeneous and stationary field.
As in \citet*{mccomas_bretherton_1977}, \citet{dong_buhler_smith_2020} and \citetalias{cox_kafiabad_vanneste_2023}, we retain time dependence in $\omf$ but later will simplify to the time-independent case.

We start with the action conservation equation \eqref{actioncons} with total frequency \eqref{totfreq}. 
Following \S A.1 of \citetalias{kafiabad_savva_vanneste_2019}, we introduce a small bookkeeping parameter $\epsilon$, making the substitution $\omf \rightarrow \epsilon \omf$ to enforce the assumption that the perturbation terms are weak enough to be dominated by the bare frequency. We define slow time and space scales $T = \epsilon^2 t$, $\bX = \epsilon^2 \bx$ and expand the action $a(\bx, \bX, \bk, t, T)$ in powers of $\epsilon$,
\begin{equation}\label{actionexpansion}
    a = a\num{0}(\bX, \bk, T) + \epsilon a\num{1}(\bx, \bX, \bk, t, T) + \dots
\end{equation}
Allowing $a\num{0}$ to vary only on slow time and space scales immediately satisfies the leading order equation. At $O(\epsilon)$,
\begin{equation}
    \pa{t} a\num{1} + c_i \pa{x_i} a\num{1} = \pa{x_i} \omf \pa{k_i} a\num{0},
\end{equation}
where $c_i$ is the $i^{\text{th}}$ component of
\begin{equation}\label{lo-gc}
    \bc = \gk \omi,
\end{equation}
the (leading order contribution to the) group velocity of the waves. This has the solution
\begin{equation}\label{order1}
    a\num{1}(\bx, \bX, \bk, t, T) = \int_0^t \pa{x_j} \omf (\bx - \bc s,\bk, t - s) \, \d s \, \pa{k_j} a\num{0}.
\end{equation}
At $O(\epsilon^2)$, we average to eliminate $a\num{2}$ terms and find
\begin{equation}\label{order2}
    \pa{T} a\num{0} + c_i \pa{X_i} a \num{0} = \langle \pa{k_i} ( a\num{1} \pa{x_i} \omf ) \rangle,
\end{equation}
where we have used spatial homogeneity. Unlike \citetalias{kafiabad_savva_vanneste_2019} and following \citet{villas_boas_young_2020}, we do not require $\pa{x_i} \pa{k_i} \omega_1 = 0$, equivalent to incompressibility for $\omega_1 = \bU \bcdot \bk$.  We substitute \eqref{order1} into \eqref{order2}, taking the upper bound of the integral to be $t \rightarrow \infty$, appropriate for slowly evolving inhomogeneities. Then
\begin{equation}
    \pa{T} a\num{0} + c_i \pa{X_i} a \num{0} = \pa{k_i} \left( \D_{ij} \pa{k_j} a\num{0} \right),
\end{equation}
where
\begin{equation}\label{diffusivity1}
    \D_{ij} = \int_0^\infty \langle \pa{x_i} \omf(\bx, \bk, t) \pa{x_j} \omf(\bx - \bc s, \bk, t - s) \rangle \, \d s,
\end{equation}
with $\langle \cdot \rangle$ the ensemble average, are the components of the diffusivity tensor $\bD$.
Setting the bookkeeping parameter to $1$ gives the diffusion equation,
\begin{equation}\label{diffusioneqn}
    \pa{t} a + \bc \bcdot \gx a = \gk \bcdot \left( \bD \bcdot \gk a\right).
\end{equation}
Considering the special case of $\omf = \bU \bcdot \bk$, the diffusivity \eqref{diffusivity1} reduces to the \citet{mccomas_bretherton_1977} diffusivity
\begin{equation}\label{KSVD}
    \D_{ij} = k_m k_n \int_0^\infty \langle \pa{x_i} U_m(\bx, t) \pa{x_j} U_n(\bx - \bc s, t - s) \rangle \, \d s.
\end{equation}


To evaluate the diffusivity, we introduce the correlation function
\begin{equation}\label{lambda}
    \Lambda(\by,\bk, r)= \langle \omf(\bx, \bk, t) \omf(\bx + \by, \bk, t + r)\rangle,
\end{equation}
and rewrite \eqref{diffusivity1} as
\begin{equation}
    \D_{ij} = -\frac{1}{2}\int_{-\infty}^\infty \frac{\partial^2 \Lambda}{\partial y_i \partial y_j}(\bc s, \bk, s) \, \d s,
\end{equation}
where we use $\Lambda(\by, \bk, r) = \Lambda(-\by, \bk, -r)$ to extend the limits of integration. Defining the Fourier transformed correlation function $\hat{\Lambda}$ through its inverse
\begin{equation}\label{inFT}
    \Lambda(\bx,\bk, t) = \int_{\mathbb{R}^{n+1}} \hat{\Lambda}(\bK,\bk, \iOmega) \e^{\i(\bK \bcdot \bx - \iOmega t)} \,\d\bK \d \iOmega,
\end{equation}
the diffusivity becomes
\begin{equation}\label{general_diff3}
    \D_{ij} = \upi \int_{\mathbb{R}^{n+1}} K_i K_j \hat{\Lambda}(\bK, \bk, \iOmega)\delta(\bK \bcdot \bc - \iOmega ) \, \d \bK \d \iOmega,
\end{equation}
where we have used that $\int_{\mathbb{R}} \exp{(\i(\bK \bcdot \bc - \iOmega )s)} \d s = 2 \upi \delta(\bK \bcdot \bc - \iOmega )$. \textcolor{black}{For} time-independent $\omf$, this reduces to
\begin{equation}\label{timeindFTdiff}
    \D_{ij} = \upi \int_{\mathbb{R}^{n}} K_i K_j \hat{\Lambda}(\bK, \bk)\delta(\bK \bcdot \bc) \, \d \bK.
\end{equation}
Next, we evaluate this expression for our two example systems and in the shallow water case, support our findings with ray tracing simulations. In the Boussinesq case, we find the forced equilibrium wave energy spectrum.



\subsection{Shallow water}\label{sec:GDSW}


To evaluate the shallow water diffusivity, we first evaluate the correlation function \eqref{lambda}, then find its Fourier transform and substitute into \eqref{general_diff3}. We introduce the background flow stream function $\psi$ such that 
\begin{equation}\label{streamfunction}
    \bU = \gx^{\perp} \psi = (-\pa{x_2} \psi, \pa{x_1} \psi, 0) \quad \textrm{and} \quad g \DH = f \psi,\addtocounter{equation}{1}\tag{\theequation a,b}
\end{equation}
where $\gx^{\perp} = (-\partial_{x_2},\partial_{x_1})$ is the skew gradient \textcolor{black}{and we use geostrophic balance \eqref{geobalSW}}. 
Substituting this and the frequency \eqref{omfSW} into \eqref{lambda}  yields
\begin{equation}\label{lambdaSW2}
    \Lambda(\by,\bk, r)= \underbrace{\vphantom{\frac{f^2 k_h^4}{4 \omi^2}}-(\bk \bcdot \gy^{\perp})^2\langle  \psi(\bx, t) \psi(\bx + \by, t + r)\rangle}_{\text{Doppler shift}}
    \, + \, \underbrace{\frac{f^2 k_h^4}{4 \omi^2} \langle \psi(\bx, t)\psi(\bx + \by, t + r)\rangle}_{\text{height fluctuation}}. 
\end{equation}
For the Doppler shift term, the skew gradients have been moved through the ensemble average using integration by parts and the symmetry of $\bx$ and $\by$ arguments. Notably, there are no cross terms in \eqref{lambdaSW2} and the height fluctuation and Doppler shift effects are uncoupled. This is because, using integration by parts under the ensemble average,
\begin{align}\label{crossterms}
    \text{cross terms} &\propto \langle  \psi(\bx + \by, t + r) \bk \bcdot \gx^{\perp} \psi(\bx, t) \rangle + \langle \psi(\bx, t) \bk \bcdot \gx^{\perp} \psi(\bx + \by, t + r) \rangle\notag\\
    &\propto - \langle \psi(\bx, t) \bk \bcdot \gx^{\perp} \psi(\bx + \by, t + r) \rangle + \langle \psi(\bx, t) \bk \bcdot \gx^{\perp} \psi(\bx + \by, t + r) \rangle\notag\\
    &= 0.
\end{align}


\textcolor{black}{The Fourier transform of the correlation function \eqref{lambdaSW2} is
\begin{equation}\label{lambdaSWFT}
    \hat{\Lambda}(\bK, \bk, \iOmega) = |\bk \times \bK|^2 E_\psi(\bK, \iOmega) \, + \,  \frac{f^2 k_h^4}{4 \omi^2} E_\psi(\bK, \iOmega)
\end{equation}
where
\begin{equation}\label{streamfuncspec}
    E_\psi(\bK, \iOmega) = \langle \hat{\psi}(-K, -\iOmega) \hat{\psi}(K, \iOmega)\rangle
\end{equation}
is the spectrum of the stream function. Defining the horizontal energy spectrum of the background flow as
\begin{equation}\label{KineticSpec}
    E(\bK, \iOmega) = K_h^2 E_\psi /2,
\end{equation}
we rewrite \eqref{lambdaSWFT} as
\begin{equation}
    \hat{\Lambda}(\bK, \bk, \iOmega) = 2 k_h^2 \sin^2 \gamma E(\bK, \iOmega)\, + \, \frac{f^2 k_h^4}{2 K_h^2 \omi^2} E(\bK, \iOmega)\label{lambdaSWFT2}
\end{equation}
where $K_h$ is the horizontal wavenumber of the background flow and $\gamma$ is the angle between $\bK$ and $\bk$.}
Combined with \eqref{general_diff3}, we have a diffusivity that accounts for height fluctuation effects for a time-dependent flow.

As proof of concept, we simplify to a time-independent flow. This 1) allows for inexpensive ray-tracing simulations in \S\ref{sec:rays}; and 2) results in a simpler form for the diffusivity.
Adding time dependence is possible, as done solely for the Doppler shift effect by \citet{dong_buhler_smith_2020} and -- for the full 3d set-up -- \citetalias{cox_kafiabad_vanneste_2023}. 
Combining the time-independent-flow limit of \eqref{lambdaSWFT2} with \eqref{timeindFTdiff}, we have
\begin{align}
    \D_{ij} = \, &2 \upi k_h^2 \int_{0}^{\infty} \d K_h\int_{0}^{2\upi} \d \gamma \, K_h  K_i K_j \sin^2 \gamma E(\bK) \delta(\bK \bcdot \bc)\notag\\ &+ \frac{\upi f^2 k_h^4}{2 \omi^2} \int_{0}^{\infty} \d K_h\int_{0}^{2\upi} \d \gamma \, \frac{K_i K_j}{K_h} E(\bK)\delta(\bK \bcdot \bc)\label{DijSW},
\end{align}
in polar coordinates $\bK = (K_h, \gamma)$. $E(\bK)$ is the flow kinetic energy spectrum marginalised over frequencies.
In the local polar basis $(\be_{k_h}, \be_{\phi})$ associated with $\bk$, this diffusivity has one non-zero component,
\begin{equation}\label{dphiphi}
    \D_{\phi \phi} = \be_{\phi} \bcdot \bD \bcdot \be_{\phi}.
\end{equation}
This is because the constant frequency surface in spectral space is the circle $\omega = \omi(k_h) + O(\epsilon) = \text{const}$. There is no diffusion of action across this circle because for a time-independent linear system, wave frequency does not change. This means there are no radial components of the diffusivity, only \eqref{dphiphi}. We see this explicitly by expanding $\bK$ as
\begin{equation}\label{Kexp}
    \bK = K_h \cos \gamma \be_{k_h} + K_h \sin \gamma \be_{\phi}.
\end{equation}
Then, noting $\delta(\bK \bcdot \bc) \propto \delta(\cos \gamma)$ as $\bc = \gk \omega_0 (k_h) = c \be_{k_h}$ shows that any integrand containing $\cos \gamma$  integrates to zero (providing the energy spectrum $E(\bK)$ is well-behaved). This is the case for all components of \eqref{DijSW} bar \eqref{dphiphi}, which, using \eqref{Kexp}, is
\begin{align}
    \D_{\phi \phi} = \, &2 \upi k_h^2 \int_{0}^{\infty} \d K_h\int_{0}^{2\upi} \d \gamma \, K_h^3 \sin^4 \gamma E(\bK) \delta(K_hc\cos\gamma)\notag\\ &+ \frac{\upi f^2 k_h^4}{2 \omi^2} \int_{0}^{\infty} \d K_h\int_{0}^{2\upi} \d \gamma \, K_h \sin^2 \gamma E(\bK)\delta(K_hc \cos \gamma)\label{DijSW2}.
\end{align}
Assuming flow isotropy such that $E(\bK) = E(K_h)$, we integrate in $\gamma$ to get
\begin{equation}
    \D_{\phi \phi} = \underbrace{\frac{4 \upi k_h^2}{c \vphantom{\omega^2_0}} \int_{0}^{\infty} K_h^2 E(K_h) \, \d K_h}_{\text{Doppler shift}} \, + \, \underbrace{\frac{\upi f^2 k_h^4}{c \omi^2} \int_{0}^{\infty} E(K_h) \, \d K_h}_{\text{height fluctuation}}\label{DijSW3}.
\end{equation}
With this single component, the diffusion equation \eqref{diffusioneqn} reduces to
\begin{equation}\label{diffeqnSW}
    \pa{t} a + \bc \bcdot \gx a = \mu \pa{\phi \phi} a,
\end{equation}
where we define the directional diffusivity
\begin{equation}\label{gradient}
    \mu = \D_{\phi \phi} / k_h^2.
\end{equation}

To examine the relative importance of each term in \eqref{DijSW3}, we consider the ratio between the two,
\begin{equation}\label{Rswprime}
    \Rsw' = \frac{\left[\D_{\phi \phi}\right]_{\text{height fluctuation}}}{\left[\D_{\phi \phi}\right]_\text{Doppler shift}} = \frac{f^2 k_h^2}{4 \omi^2} \frac{\int_{0}^{\infty} E(K_h) \, \d K_h}{\int_{0}^{\infty} K_h^2 E(K_h) \, \d K_h} \sim \frac{(k_h/K_*)^2}{4 (1 + Bu \, (k_h/K_*)^2)}.
\end{equation}
\textcolor{black}{Here, we approximate the integrals by
\begin{equation}\label{USTAR}
    2 \upi \int_{0}^{\infty} K_h^2 E(K_h) \, \d K_h \sim K_* U_*^2/2 \quad \text{and} \quad 2 \upi \int_{0}^{\infty} E(K_h) \, \d K_h \sim U_*^2/(2 K_*) \addtocounter{equation}{1}\tag{\theequation a,b}
\end{equation}
using the characteristic speed and horizontal wavenumber of the flow.
} The final expression in \eqref{Rswprime} is the square of the scaling argument estimate $\Rsw$ \eqref{ratioSW} which comes from the wave dispersion relation. This is sensible -- the diffusivity \eqref{diffusivity1} consists of a square frequency term. This means that in the diffusion regime, the scaling argument of \S\ref{sec:scale_sw} is an underestimate for $\Rsw > 1$ and overestimate for $\Rsw < 1$. In figure \ref{fig:ratio2}, the adjustment from $\Rsw$ to $\Rsw'$ is shown.


\subsection{Shallow water ray tracing}\label{sec:rays}
The 2D diffusion equation \eqref{diffeqnSW} has an exact solution as outlined in \S4 of \citet{villas_boas_young_2020}. We use this solution and ray tracing -- a numerical method to find constant-wave-action trajectories -- to assess the validity of the diffusion approximation and form of the diffusivity \eqref{DijSW3}.





We start by multiplying the diffusion equation by $\cos \phi$ and integrating over the entire $(x, y)$-plane and $\phi \in (-\upi, \upi)$ to get
\begin{equation}
    \frac{\d}{\d t} \iiint \cos \phi \, a \, \d x \d y \d \phi = -\mu\iiint \cos \phi\,  a \, \d x \d y \d \phi.
\end{equation}
Note that without pre-multiplying by $\cos \phi$, the left-hand side is zero by action conservation. 
Here, the $\bc \bcdot \gx a$ term disappears under the $(x,y)$ integrals because the $a$ we consider is ensemble averaged and we assume statistical spatial homogeneity in deriving the diffusion equation. Integrating with respect to time yields
\begin{equation}\label{full_soln}
    \ln{(\langle \cos \phi \rangle_a)} = - \mu t + \ln{(\langle \cos \phi_0 \rangle_a)},
\end{equation}
where $\langle \cdot \rangle_a$ denotes the action-weighted ensemble average and $\phi_0$ is the initial angle. This is an alternative form of (4.3) in \citet{villas_boas_young_2020}.

We confirm that \eqref{full_soln} holds by ray tracing: solving the characteristic equations of the action conservation equation \eqref{actioncons} numerically to give many constant-action trajectories and taking an ensemble average. These ray equations are
\begin{equation}
    \partial_t \bx = \gk \omega, \quad \text{and} \quad
    \partial_t \bk = -\gx \omega. \addtocounter{equation}{1}\tag{\theequation a,b}\label{chars}
\end{equation}
We define non-dimensional quantities
\begin{equation}\label{nondim}
    t' = t f, \quad \bk' = \bk/K_*, \quad \bx' = \bx K_*, \quad \text{and} \quad \bU' = \bU/U_*, \addtocounter{equation}{1}\tag{\theequation a--d}  
\end{equation}
(and $\psi' = \psi K_*/U_*$). Substituting the approximate frequency for shallow water waves \eqref{SWtot_2} into \eqref{chars} gives
\begin{subequations}
    \begin{align}
    &\partial_{t'} \bx' = &  &\frac{Bu}{(1 + Bu k_h'^2)^{1/2}}\bk' & &+ &  & Ro \bU'    &   &+ &  & Ro \frac{2 + Bu k_h'^2}{2 (1 + Bu k_h'^2)^{3/2}} \psi' \bk'  , \label{charx}\\
    &\partial_{t'} \bk' = &  &0    &   &- &  &Ro \boldsymbol{\nabla}_{\bx'}\left(\bU' \bcdot \bk'\right) &  &- &  &Ro \frac{k_h'^2}{2 (1 + Bu k_h'^2) }\bU'^{\perp},\label{chark}\\& & 
    &\underbrace{\hphantom{\frac{Bu}{(1 + Bu k_h'^2)^{1/2}}\bk'}}_{\text{no flow}} & & & &\underbrace{\hphantom{Ro \boldsymbol{\nabla}_{\bx'}\left(\bU' \bcdot \bk'\right)}}_{\text{Doppler shift}} & & & &\underbrace{\hphantom{Ro \frac{2 + Bu k_h'^2}{2 (1 + Bu k_h'^2)^{3/2}} \psi' \bk'}}_{\text{height fluctuation}}\notag
    \end{align}
\end{subequations}
where
\begin{equation}\label{rossbyeqn}
    Ro = \frac{U_*K_*}{f}
\end{equation}
is the flow Rossby number and we introduce $\bU^\perp = (V, -U, 0)$.

\begin{figure}
    \centering
    \includegraphics{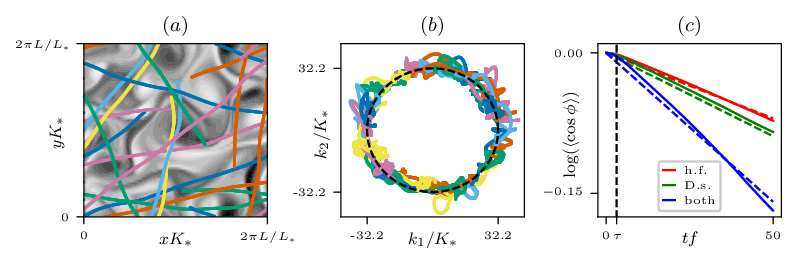}
    \caption{Ray trajectories satisfying the characteristic equations \eqref{charx}--\eqref{chark}, propagating in the flow described in \S\ref{sec:rays} with $Ro = 0.03$, $Bu = 0.32$ and $K_* = 3.10$. Panel (a) shows a sample of 10 rays in physical space superimposed on the flow vorticity field (darker sections indicate higher-magnitude vorticity). Panel (b) shows a sample of 50 rays in spectral space initialised on a constant-frequency circle given by the dotted black line. Panel (c) gives the exact solution \eqref{full_soln} approximated by the rays' ensemble average (solid lines) and the diffusivity \eqref{DijSW3} (dashed lines). Red lines are calculated solely with the height fluctuation terms in \eqref{DijSW3} and \eqref{charx}--\eqref{chark}; green lines are solely Doppler shift and; blue lines are the entire system with both Doppler and height fluctuation effects. The vertical dashed line indicates $\tau$, the point at which the gradient calculation for figure \ref{fig:robu} begins.}
    \label{fig:rays}
\end{figure}

The rays propagate in a flow which is a snapshot of a fully-evolved 2D quasi-geostrophic Navier-Stokes simulation, solved
using a pseudo-spectral method and Crank-Nicholson time-stepping with time step $\d t = 0.01$.
The doubly periodic domain \textcolor{black}{$\bx \in [0, L]$ is discretised by $1024^2$ gridpoints.} 
Viscosity is $\nu = 10^{-5}$. The initial wavenumber distribution of the flow is Gaussian. We prescribe 9 values of the Rossby radius of deformation $L_D = Bu^{1/2}/K_*$ and 3 values of the initial wavenumber $K_*$ to give $3^3$ flows. This gives final snapshots with varying $Ro$ and $Bu$.
A typical flow is shown in panel (a) of figure \ref{fig:rays}.

We perform ray simulations with the inhomogeneity $\omf$ given by (a) Doppler shift only; (b) height fluctuation only and; (c) both Doppler shift and height fluctuation.

We initialise $50^2$ rays equally distributed across the periodic $\bx$-domain with a constant horizontal wavenumber $k_h \approx 32.2 K_*$ (within the WKB regime) and random initial angle. 
By giving the rays different initial angles, our rays sample more of the flow and the ensemble average approaches the exact solution more rapidly as the number of rays increases than it would with a constant initial angle. For simplicity, we consider the angle $\phi$ to be the angle of deviation from the ray's initial angle meaning $\ln{(\langle \cos \phi_0 \rangle_a)} = 0$. This is possible because of the flow's statistical isotropy. The angle of deviation for each ray is calculated at each time step, and $\cos \phi$ is averaged over. We choose action $a=1$ for each ray so that $\langle \cdot \rangle_a$ and the ensemble average are identical. Some of the $3^3$ flows have very similar Burger or Rossby numbers;  we average over these Burger and Rossby numbers to leave $3^2$ results with distinct Burger and Rossby numbers.
We estimate $\mu$, \eqref{gradient}, the negative gradient of $\ln{(\langle \cos \phi \rangle_a)}$ against time $t$ with $\D_{\phi \phi}$ computed from \eqref{DijSW3}. We compare this to the simulation gradient. 

An example ray tracing simulation is shown in figure \ref{fig:rays} in physical and wavevector space. The Rossby number of the flow is the small parameter in \S\ref{sec:GD}, $\epsilon \sim Ro \sim O(0.01)$, and so the flow is weak and the rays are only slightly deflected from their initial angle of propagation. In $\bk$ space, the small Rossby number characterises the thickness of the constant-frequency ring. The exact solution and simulation approximation of \eqref{full_soln} are shown in figure \ref{fig:rays}(c). This confirms the validity of the diffusion approximation and of the formulas \eqref{DijSW3}. As expected from these formulas, the height fluctuations and Doppler shift make comparable contributions to the spectral diffusivity.

\begin{figure}
    \centering
    \includegraphics{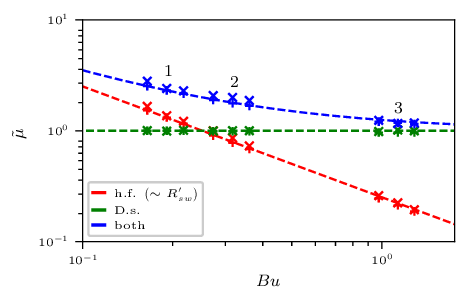}
    \caption{Non-dimensional directional diffusivity $\tilde{\mu}$, as defined by \eqref{gradient} and \eqref{mutilde}, plotted against flow Burger number $Bu$. Red lines correspond to the height fluctuation component of the directional diffusivity, green lines to the Doppler shift component and blue lines to the full directional diffusivity. Crosses indicate the estimates of $\tilde{\mu}$ from the $50^2$-ray simulations and exact solution \eqref{full_soln}, pluses are computed from the diffusivity expression \eqref{DijSW3} and dotted lines are the power laws predicted by \eqref{mutilde}, fitted to the first of the exact data points. The 9 different Burger numbers cover 3 sets of Rossby number, labelled $1$ ($Ro \approx 0.023$), $2$ ($Ro \approx 0.030$) and $3$ ($Ro \approx 0.057$). The Doppler shift directional diffusivity $\tilde{\mu} \approx 1$ and so the height fluctuation directional diffusivity corresponds to the ratio between the two, $\Rsw'$ \eqref{Rswprime}.}
    \label{fig:robu}
\end{figure}

The $Ro$ and $Bu$ dependence of $\mu$ can be seen from \eqref{DijSW3}. We define the non-dimensional directional diffusivity $\tilde{\mu}$ as
\begin{equation}\label{mutilde}
    \tilde{\mu} = \frac{Bu^{1/2 }}{f Ro^2}\mu \sim \underbrace{\frac{(\omi/f)}{Bu^{1/2}  (k_h/K_*)}}_{\text{Doppler shift}} \, + \, \underbrace{\frac{ (k_h/K_*) }{4 Bu^{1/2} (\omi/f)}}_{\text{height fluctuation}}
    \rightarrow \underbrace{\vphantom{\frac{(\omi/f)}{Bu^{1/2}  (k_h/K_*)}}1}_{\text{D.s.}} \,+\, \underbrace{\vphantom{\frac{(\omi/f)}{Bu^{1/2}  (k_h/K_*)}}(4Bu)^{-1}}_{\text{h.f.}},
\end{equation}
\textcolor{black}{where the final limit holds for $(k_h/K_*)^2 \gg 1$ and we have used \eqref{USTAR}. The scaling of $\mu$ is chosen so that the Doppler-shift component of $\tilde{\mu}$ is independent of $Bu$ and $Ro$.}
 Note the ratio between height fluctuation and Doppler shift effects is in agreement with $\Rsw'$, \eqref{Rswprime}. In figure \ref{fig:robu}, $\tilde{\mu}$ is displayed for a range of Burger numbers. The Burger dependence of \eqref{mutilde} is confirmed for each contribution and, because 3 separate Rossby numbers are used in the plot, so too is the Rossby dependence.

The characteristic flow velocity $U_*$ is chosen such that the first expression of \eqref{USTAR} holds exactly and $\tilde{\mu} = 1$ is the exact limit of the non-dimensionalised, Doppler shift directional diffusivity. Then the height fluctuation $\tilde{\mu}$ shown in figure \ref{fig:robu} coincides with the ratio between the height fluctuation and Doppler shift effects $\Rsw'$. We see good agreement between the height fluctuation $\tilde{\mu}$ computed exactly and from simulations and thus confirm our estimate of $\Rsw'$ given by \eqref{Rswprime}.



In deriving the diffusivity \eqref{diffusivity1}, we approximate an integral between $0$ and $t$ by one between $0$ and $\infty$ 
and so there is a delay between the start of the ray simulation to the point at which the exact solution \eqref{full_soln} is well approximated.
Therefore, our $\mu$ estimates in figure \ref{fig:robu} begin a short time $\tau/f$ after the simulations have begun, as indicated in figure \ref{fig:rays}(c). 

We stress that the good agreement in figures \ref{fig:rays}(c) and \ref{fig:robu} between the $\mu$ \eqref{gradient} computed from the diffusivity \eqref{DijSW3} and the $\mu$ of \eqref{full_soln} estimated from the ensemble average of rays validates the diffusion approximation \eqref{diffeqnSW} of the action conservation equation \eqref{actioncons} with inhomogeneities of the form \eqref{omfSW}.

\subsection{3D Boussinesq}\label{sec:GDB}

The method to evaluate the general diffusivity \eqref{general_diff3} for a Boussinesq system with vertical buoyancy gradients is similar to the shallow water system. We give the result here, providing a full derivation in appendix \ref{app:boussdiff}.
For a Doppler-shift-induced diffusivity, \citetalias{cox_kafiabad_vanneste_2023} show that it is well justified to assume a slowly evolving flow is time independent. The same likely applies to a buoyancy-gradient-induced diffusivity, so we focus on the time-independent case.


The two non-zero components of the diffusivities for a Boussinesq system with vertical buoyancy gradients and dispersion relation \eqref{omtotBous2} are
\begin{align}\label{DkkBouss}
    \D_{kk} = &\frac{4\upi k^3 \omi \sin^2 \theta}{(\Nb^2 - f^2) |\cos^5 \theta|} \int_{0}^{\infty} \left. \int_{\theta}^{\upi - \theta} K^3 \cos^2 \iTheta (\cot^2 \theta - \cot^2 \iTheta)^{1/2} E(K, \iTheta) \, \d K \d \iTheta \right\} \text{\small Doppler shift}\notag\\
    &+ \frac{\upi k f^2 \sin^2 \theta }{ \omi (\Nb^2 - f^2)|\cos^3 \theta|}  \left. \int_{0}^{\infty} \int_{\theta}^{\upi - \theta} \frac{K^5  \cos^6 \iTheta E(K, \iTheta)}{ \sin^2 \iTheta (\cot^2 \theta - \cot^2 \iTheta)^{1/2}} \, \d K \d \iTheta\right\} \text{\small buoyancy fluctuation},
\end{align}
and
\begin{align}\label{DphiphiBouss}
    \D_{\phi \phi} = &\frac{4\upi k^3 \omi \sin^4 \theta}{(\Nb^2 - f^2) |\cos^5 \theta|} \left. \int_{0}^{\infty} \int_{\theta}^{\upi - \theta} K^3 \sin^2 \Theta  (\cot^2 \theta - \cot^2 \iTheta)^{3/2}E(K, \iTheta)\, \d K \d \iTheta\right\} \text{\small Doppler shift}\notag\\
    &+ \frac{\upi k f^2 \sin^4 \theta }{ \omi (\Nb^2 - f^2) |\cos^3 \theta|}  \left. \int_{0}^{\infty} \int_{\theta}^{\upi - \theta}K^5 \cos^4 \iTheta (\cot^2 \theta - \cot^2 \iTheta)^{1/2} E(K, \iTheta)\, \d K \d \iTheta\right\}\text{\small buoy.\ fluct.},
\end{align}
where we use polar coordinates $\bK = (K, \gamma, \iTheta)$ for the background flow and $E$ is the flow kinetic energy spectrum, as defined in the shallow water case 
\textcolor{black}{\eqref{streamfuncspec}--\eqref{KineticSpec}}.
Here, $\gamma = \mathit{\Phi} - \phi$ is the angle between the horizontal components of $\bK$ and $\bk$, \textcolor{black}{with $\mathit{\Phi}$ the azimuthal angle of $\bK$}. The Doppler shift and buoyancy fluctuation diffusivities are uncoupled, as with the inhomogeneities in the shallow water system. This is true of a time-dependent background flow also (see appendix \ref{app:boussdiff}).

Only the kinetic energy spectrum of the flow appears in \eqref{DkkBouss}--\eqref{DphiphiBouss} because we assume thermal wind balance. More generally, the diffusivity will be expressed in terms of both potential and kinetic energy spectra.

The $(\cot^2\theta -\cot^2\iTheta)^{-1/2}$ factor in the integrand of the buoyancy fluctuation term of $\D_{kk}$ behaves like $\delta^{-1/2}$ a small distance $\delta$ from the singularities at $\iTheta = \theta, \pi-\theta$. Therefore, it is integrable and does not cause the diffusivity to diverge.

\begin{figure}
    \centering
    \includegraphics[width=.75\textwidth]{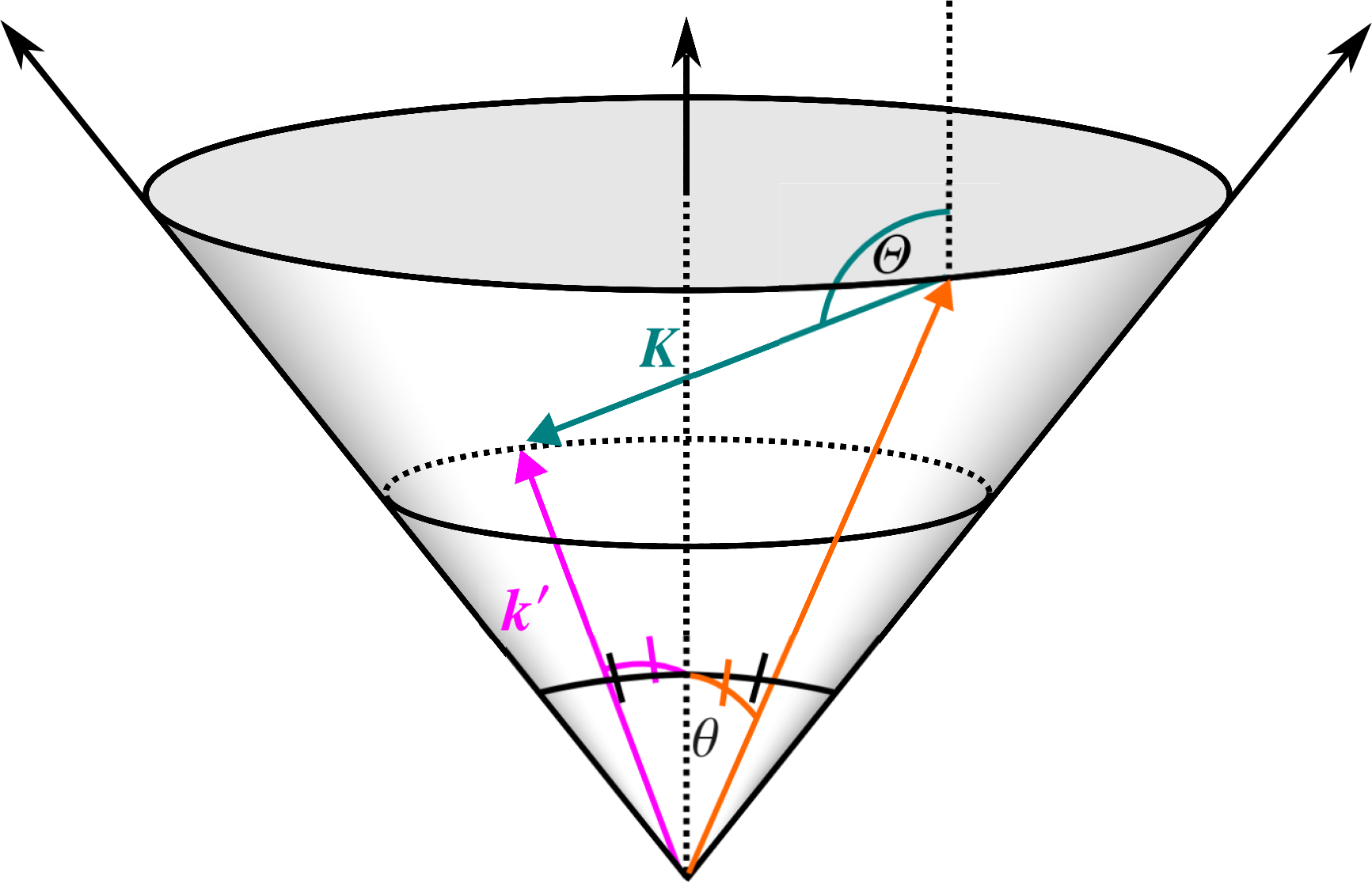}
    \caption{Schematic of the scattering interaction between an incident wave $\bk$ and the background flow $\bK$ resulting in a scattered wave $\bk'$. The two waves have the same frequency $\omi(\theta)$ and thus the background flow connects two points on the constant-frequency cone. For an arbitrary scattering interaction, angle $\iTheta$ of the flow's wavevector is bounded between $\theta$ and $\upi - \theta$ i.e.\ the angular range of a vector between any two points on the cone.}
    \label{fig:icecreamcone}
\end{figure}

Only the part of the flow spectrum with wavevectors of polar angle $\iTheta \in (\theta, \pi - \theta)$ contributes to the diffusivity integrals. IGW diffusion is a sub-regime of IGW scattering. This triadic interaction occurs between an incident wave $\bk$ and the background flow $\bK$, resulting in a scattered wave $\bk'$ of the same frequency. In wavevector space, $\bk + \bK = \bk'$. As the incident wave and scattered wave are of the same frequency, $\bK$ connects two points on the cone of constant frequency. Thus, the range of the $\bK$'s polar angle $\iTheta$ is the range of angles connecting any two points on the cone i.e.\ $(\theta, \upi - \theta)$. This is demonstrated in figure \ref{fig:icecreamcone}. Significantly, waves of frequency $\omi(\theta)$ will not be scattered by flow fluctuations with wavevectors outside of these angle bounds, regardless of the flow's energy. 

These invisible flow regions make it difficult to apply the scaling argument of \S\ref{sec:scalingarg_bouss}. The polar angle of the characteristic wavevector of the flow $\bK_*$ does not necessarily lie within $(\theta, \pi-\theta)$, in which case $\RB$ \eqref{ratioB} computed at $\bK_*$ is not meaningful. Instead, a dominant wavevector -- the characteristic wavevector of the portion of the flow which scatters the waves -- must be used.

We define $\RB'$ as the ratio between the buoyancy fluctuation and Doppler shift diffusivites for radial and azimuthal components,
\begin{equation}\label{rbprimes}
    {\RB'}_{,kk} = \frac{[\D_{kk}]_{\text{buoy.\ fluct.}}}{[\D_{kk}]_{\text{Doppler shift}}} \quad \text{and} \quad {\RB'}_{,\phi \phi} = \frac{[\D_{\phi \phi}]_{\text{buoy.\ fluct.}}}{[\D_{\phi \phi}]_{\text{Doppler shift}}}.\addtocounter{equation}{1}\tag{\theequation a,b}
\end{equation}
Both ratios scale like $(k_h/K_*)^{-2}$ through the diffusivity prefactors' $k$ dependence. The frequency behaviour is more complicated, requiring the diffusivity integrals to be evaluated. In appendix \ref{app:RBpowers}, we show that $\RBpk \rightarrow 0$ for $\omi/f \rightarrow 1$ and for larger frequencies and high aspect ratios $\alpha$ \eqref{aspect}, $\RBpk \sim (\omi/f)^{-2}$. This means that for waves in the WKB regime with high aspect ratios and frequencies, vertical buoyancy gradients can be neglected.

\begin{figure}
  \centerline{\includegraphics{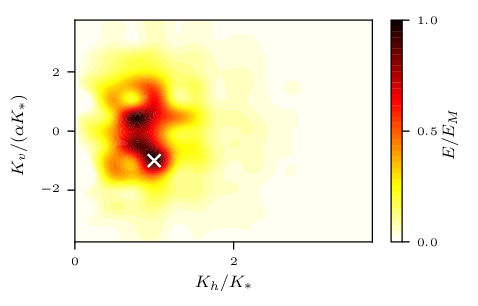}}
  \caption{The geostrophic flow component $E$, smoothed and scaled by a maximum value $E_M$, of the full Boussinesq simulation in \citetalias{cox_kafiabad_vanneste_2023}. The horizontal and vertical flow wavenumbers ($K_h, K_v$) are scaled by the characteristic wavevector $(K_h, K_v) = (K_*, \alpha K_*)$ -- with $\alpha$ the aspect ratio of the flow \eqref{aspect} -- marked by a white cross, at which the geostrophic energy is maximum.}
\label{fig:geoflow}
\end{figure}

We compute the ratio of diffusivity components for a typical quasi-geostrophic flow and find good agreement with the $-2$ power laws.
The geostrophic energy spectrum used is extracted from a snapshot of the full Boussinesq simulation described in \citetalias{cox_kafiabad_vanneste_2023} and is pictured in figure \ref{fig:geoflow}. For this spectrum, $N/f = 32.0$ and the aspect ratio, \eqref{aspect}, $\alpha \approx 16.0$.

\begin{figure}
  \centerline{\includegraphics{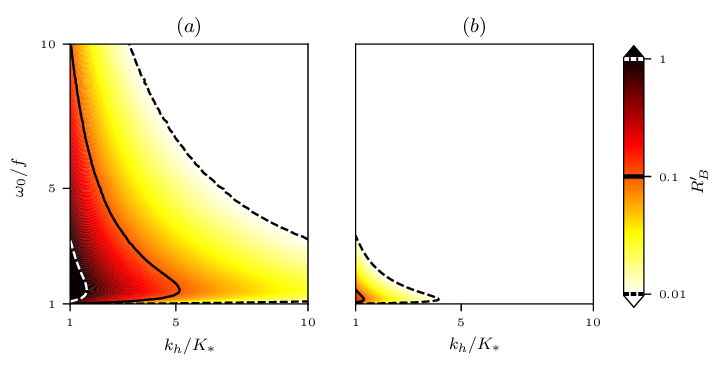}}
  \caption{Radial and azimuthal ratios $(a)$ $\RBpk$ and $(b)$ $\RBpp$ as given in \eqref{rbprimes}, the ratio of buoyancy fluctuation and Doppler shift diffusivities for a snapshot of the full Boussinesq simulation of \citetalias{cox_kafiabad_vanneste_2023}, against non-dimensionalised frequency $\omega_0/f$ and horizontal wavenumber $k_h/K_*$. Contours are shown for $\RB\textcolor{black}{'} = 0.01$ (dotted black), $0.1$ (solid black) and $1$ (dotted white).}
\label{fig:ratioBfilter}
\end{figure}
\begin{figure}
    \centering
    \includegraphics{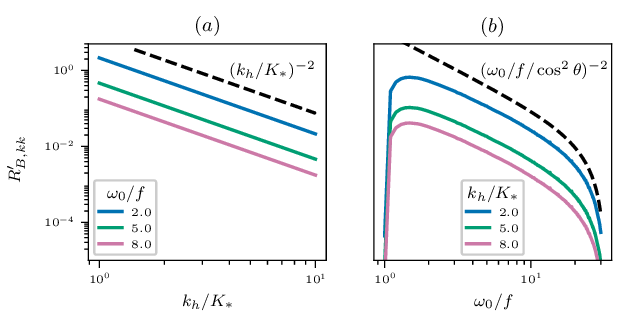}
    \caption{Cross sections of radial ratio $\RBpk$ \eqref{rbprimes}, as shown in figure \ref{fig:ratioBfilter}, for constant $(a)$ $\omi/f$ and $(b)$ $k_h/K_*$. The $(\omi/f/\cos^2\theta)^{-2}$ prediction of appendix \ref{app:RBpowers} is, for small $\theta$, an $(\omi/f)^{-2}$ power law.}
    \label{fig:slices}
\end{figure}

Both ratios of diffusivities are shown in figure \ref{fig:ratioBfilter}, computed with the energy spectrum of figure \ref{fig:geoflow}. In figure \ref{fig:slices}, for the radial diffusivity ratio,
we plot cross sections of constant frequency and horizontal wavenumber and find good agreement with the $-2$ power laws.

The values of ${\RB}_{,kk}$ and ${\RB}_{,\phi\phi}$ are, for 
the WKB regime of $k_h/K_* \gg 1$, $\lesssim 0.1$. Thus, at least for the waves in \citetalias{cox_kafiabad_vanneste_2023}, we predict a weak vertical buoyancy gradient induces negligible spectral diffusion in the WKB regime.


Figure \ref{fig:ratioBfilter}, a comparison of diffusivity magnitudes, is markedly different to the comparison of dispersion relation terms, figure \ref{fig:ratioBcontour}. Although both predict a decrease in the buoyancy fluctuation effect as horizontal wavenumber increases, the scaling argument of \S\ref{sec:scalingarg_bouss} predicts the effect dominates for higher frequencies. The $(\omi/f)^{-2}$ power law predicted from diffusivities and confirmed for an example spectrum in figure \ref{fig:slices} contradicts this. We attribute this conflicting prediction to the large part of the flow spectrum not contributing to the diffusivity at higher wave frequencies. This makes the scaling arguments of \S\ref{sec:scalingarg_bouss} unreliable and calls for the exact evaluation of the diffusivities \eqref{DkkBouss}--\eqref{DphiphiBouss}.

\subsection{Forced equilibrium spectrum}\label{sec:f-e-spec}

\begin{figure}
    \centering
    \includegraphics{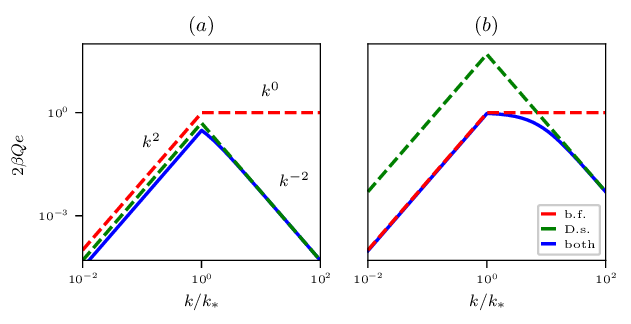}
    \caption{Forced equilibrium spectra \eqref{Generalpiecewise2}--\eqref{buoyancylimit} against non-dimensionalised wavenumber for two different values of $\beta/k_*^2 = P/(Qk_*^2)$ \eqref{Qeqn}, the $k/k_*$-independent ratio of buoyancy-induced and Doppler-shift-induced diffusivities. (a) $\beta/k_*^2 = 1$ corresponds to a diffusivity with equal contributions from buoyancy fluctuations and Doppler shift and; (b) $\beta/k_*^2 = 100$ corresponds to a buoyancy-fluctuation-dominated diffusivity. Diffusion by both effects is given in blue \eqref{Generalpiecewise2}, whilst red and green lines are spectra derived from only the buoyancy and Doppler shift terms respectively, given by \eqref{buoyancylimit} and \eqref{Dopplerlimit}.}
    \label{fig:equil}
\end{figure}

We solve the diffusion equation \eqref{diffusioneqn} exactly in the time-independent case corresponding to the equilibrium spectrum obtained under constant forcing. Our aim is to assess how the $k^{\pm 2}$ power law spectra obtained by \citetalias{kafiabad_savva_vanneste_2019} when diffusion is solely caused by Doppler shift is modified when accounting for vertical buoyancy gradients.
As in \citetalias{kafiabad_savva_vanneste_2019}, we focus on radial diffusion, assuming wave statistics independent of $\phi$ such that $\pa{\phi} a = 0$. To concisely contrast this work to theirs, we consider only the forced, equilibrium solution to \eqref{diffusioneqn}. We consider the energy density for ease of interpretation, defined as $e(k;\theta) = 2\upi k^2 \sin \theta \omega a(k;\theta)$, such that $e \, \text{d}k \text{d}\theta$ is the energy contained in the box $[k, k + \text{d}k]$ and $[\theta, \theta + \text{d} \theta]$. Thus, ignoring unimportant prefactors on the right-hand side, \eqref{diffusioneqn} becomes
\begin{equation}
    \pa{k} \left(k^2 \D_{k k} \pa{k} \frac{e}{k^2} \right) = - \delta(k - k_*).
\end{equation}
The forcing in a circle at $k = k_*$ can be generalised via integration. The minus sign ensures a positive energy spectrum. Defining
\begin{equation}\label{Qeqn}
    Q = k^3[\D_{kk}]_{\text{Doppler shift}}  \quad \text{and} \quad P = k[\D_{kk}]_{\text{buoyancy fluctuation}} 
\end{equation}
as the $k$-independent parts of the Doppler shift and buoyancy fluctuation diffusivities, this simplifies to
\begin{equation}\label{inhomo}
   Q\pa{k} \left(\left( k^5 + \beta k^3\right) \pa{k} \frac{e}{k^2} \right) = - \delta(k - k_*),
\end{equation}
where $\beta = P/Q$. This equation has the piece-wise solution found, for example, by partial fractions,
\begin{align}\label{Generalpiecewise}
    e(k) =\left\{
    \begin{array}{ll}
    A\left(1 - \frac{k^2}{\beta} \ln{\left(1 + \frac{\beta}{k^2}\right)}\right)  + B k^2 & \textrm{for}\ 0 < k < k_* \\
    C\left(1 - \frac{k^2}{\beta} \ln{\left(1 + \frac{\beta}{k^2}\right)}\right)  + D k^2  & \textrm{for} \  k > k_*
    \end{array} \right..    
\end{align}
We require $e(k)$ is bounded as $k \rightarrow \infty$ which means $D = 0$. We require zero energy at $k=0$, therefore $A = 0$. Continuity at $k = k_*$ gives $B = C(1/k_*^2 - \ln{(1 + \beta/k_*^2)}/\beta)$. The jump condition $[Q\left(k^5 + \beta k^3\right) \pa{k} \frac{e}{k^2}]_{k_*^-}^{k_*^+} = -1$ gives $C = 1/(2\beta Q)$. Thus,
\begin{align}\label{Generalpiecewise2}
    e(k) =\frac{1}{2 \beta Q}\left\{
    \begin{array}{ll}
    \frac{k^2}{k_*^2} \left(1 - \frac{k_*^2}{\beta} \ln{\left(1 + \frac{\beta}{k_*^2}\right)}\right)& \textrm{for}\ 0 < k < k_* \\
    1 - \frac{k^2}{\beta} \ln{\left(1 + \frac{\beta}{k^2}\right)} & \textrm{for} \  k > k_*
    \end{array} \right..    
\end{align}
Note that $Q \rightarrow 0$ is an artificial singularity introduced by the choice of factorisation and forcing in \eqref{inhomo}. The Doppler shift limit of $\beta \rightarrow 0$ is
\begin{align}\label{Dopplerlimit}
    e(k) =\frac{1}{4 Q k_*^2}\left\{
    \begin{array}{ll}
    (k/k_*)^2 & \textrm{for}\ 0 < k < k_* \\
    (k_*/k)^2 & \textrm{for} \  k > k_*
    \end{array} \right.,  
\end{align}
which is exactly (4.1) of \citetalias{kafiabad_savva_vanneste_2019}. The buoyancy fluctuation limit of $\beta \rightarrow \infty$ is
\begin{align}\label{buoyancylimit}
    e(k) =\frac{1}{2 P}\left\{
    \begin{array}{ll}
    (k/k_*)^2 & \textrm{for}\ 0 < k < k_* \\
    1 & \textrm{for} \  k > k_* 
    \end{array} \right..    
\end{align}
In figure \ref{fig:equil}, \eqref{Generalpiecewise2}--\eqref{buoyancylimit} are displayed for two values of $\beta/k_*^2$.

The finite energy at $k \rightarrow \infty$ of \eqref{buoyancylimit} is unphysical. However, this solution is unstable in the sense that a vanishingly small Doppler shift contribution will result in $e(k) \rightarrow 0$ as $k\rightarrow \infty$. This is because the limit of \eqref{Generalpiecewise2} as $k \rightarrow \infty$, $\beta = o(k)$ is $e(k) \rightarrow 1/(4Qk^2)$ i.e.\ the Doppler shift limit \eqref{Dopplerlimit} for $k>k_*$. This can be seen in figure \ref{fig:equil} (b). If a buoyancy fluctuation is present, so too is the Doppler shift term by the thermal wind balance \eqref{thermalwind} which means that the $k>k_*$ limit of \eqref{buoyancylimit} never occurs and the energy spectrum will always decay for large $k$ under Doppler shift.


Figure \ref{fig:equil} shows how buoyancy fluctuations affect the spectrum amplitude, mainly for $k<k_*$. For a Doppler-shift-dominated diffusivity ($\beta \rightarrow 0$), this amplitude change is negligible. For $k > k_*$, buoyancy fluctuations shallow the Doppler-shift-induced energy spectrum of \citetalias{kafiabad_savva_vanneste_2019} for a small range of intermediate wavenumbers.

The scaling argument of \S\ref{sec:scalingarg_bouss} and the ratio of radial diffusivity \eqref{DkkBouss} predict the buoyancy-induced diffusivity decays to zero for large $k/K_*$ and thus the effect of vertical buoyancy gradients is then small. Our findings in this section compound this prediction because 1) a small buoyancy-induced diffusivity has negligible impact on the wave energy spectrum for any wavenumber and; 2) any buoyancy-induced diffusivity has negligible impact on the wave energy spectrum for large $k_h/K_*$.

\section{\textcolor{black}{Diffusion induced by bottom topography}}\label{sec:topography}

In this section, we evaluate the general diffusivity of \S\ref{sec:GD} for two systems. We extend the rotating shallow water diffusivity in \S\ref{sec:GDSW} to include bottom topography (\ref{app:RSWscatt}) and evaluate the general diffusivity for surface gravity waves scattered by bottom topography and a background current (\ref{app:SWscatt}). These diffusivities are evidence that the general diffusion framework can be applied to systems beyond those already discussed and we leave further exploration (ray tracing, comparison to observation etc.) to future work.


\subsection{Rotating shallow water system}\label{app:RSWscatt}

\begin{figure}
    \centerline{\includegraphics{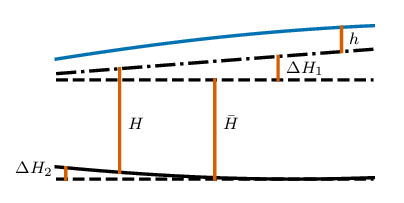}}
    \caption{Shallow water set-up with bottom topography, a modified version of figure \ref{fig:SWpaper}(b). The free surface is the solid blue line, the bottom is the solid black line and the horizontal dashed black lines enclose the constant mean height $\bar{H}$. The wave perturbations are given by $h$ and the fluctuation in height due to geostrophy, the dashed-dotted line, is given by $\DH_1$. The topographic variation in height is $\DH_2$, which is a negative quantity at the labelled location in this figure.}
    \label{fig:topog}
\end{figure}

We relax the assumption of a flat bottom made in sections \ref{sec:scale_sw}, \ref{sec:GDSW} and \ref{sec:rays} by writing
\begin{equation}\label{DH12}
    \Delta H = \Delta H_1 + \Delta H_2,
\end{equation}
where $\Delta H_1$ is induced by geostrophy and satisfies \eqref{geobalSW} and $\Delta H_2$ is the fluctuation to mean height due to bottom topography. This is shown in figure \ref{fig:topog}.



As with the height fluctuation induced by geostrophy, we assume topography varies over longer length scales than the waves so that the WKB assumption holds and action is conserved (the WKB analysis of appendix \ref{app:wkb} does not rely on the flat-bottom assumption). The amplitude of topographic variation is small relative to the mean height so that the weak interaction assumption of statistical scattering is satisfied. The perturbation \eqref{omfSW} to the bare frequency has an additional term induced by topography,
\begin{equation}\label{omfSWtop}
    \omf = \underbrace{\frac{g \DH_1 k_h^2}{2 \omi}}_{\text{geostrophic height fluctuation}} + \underbrace{\frac{g \DH_2 k_h^2}{2 \omi}}_{\text{topography}} + \underbrace{\vphantom{\frac{g \DH k_h^2}{2 \omi}} \bU \bcdot \bk}_{\text{Doppler shift}}.
\end{equation}

If the background flow and topography are uncorrelated, the diffusivity \eqref{DijSW3} has an additional term
\begin{equation}
    \D_{\phi \phi} = \underbrace{\frac{4 \upi k_h^2}{c \vphantom{\omega^2_0}} \int_{0}^{\infty} K_h^2 E(K_h) \, \d K_h}_{\text{Doppler shift}} \, + \, \underbrace{\frac{\upi f^2 k_h^4}{c \omi^2} \int_{0}^{\infty} E(K_h) \, \d K_h}_{\text{geostrophic height fluctuation}} + \, \underbrace{\frac{\upi g^2 k_h^4}{2 c \omi^2} \int_{0}^{\infty} K_h^2\mathcal{B}(K_h) \, \d K_h}_{\text{topography}}\label{DijSW3top},
\end{equation}
where $\mathcal{B}$ is the bottom topography spectrum
\begin{equation}\label{topspec}
    \mathcal{B}(\bK) = \langle \widehat{\DH_2}(-\bK) \widehat{\DH_2}(\bK)\rangle
\end{equation}
which we assume to be horizontally isotropic such that $\mathcal{B}(\bK) = \mathcal{B}(K_h)$.
The topography term in \eqref{DijSW3top} is derived by replacing $E(K_h)$ in the height fluctuation term of \eqref{DijSW3} with $ (gK_h/f)^2 \mathcal{B}(K_h)/2$, which comes from expressing $E(K_h)$ as a function of $\DH_1$ using geostrophic balance \eqref{geobalSW}, then swapping $\DH_1$ for $\DH_2$.

Should the background flow and topography be correlated, the diffusivity will contain geostrophic height fluctuation-topography and Doppler shift-topography cross terms.





\subsection{Surface waves}\label{app:SWscatt}

\citet{BrethertonGarrett68} prove that wave action is conserved for a wide class of physical systems. One of the examples they consider is surface gravity waves propagating on a mean flow with negligible rotation effects. In this case, the absolute frequency of the waves is
\begin{equation}\label{disp_surf}
    \omega = \left(g k_h \tanh{(H k_h )} \right)^{1/2} + \bU \bcdot \bk
\end{equation}
with $\bU = (U, V, 0)$ the background surface velocity.
As in figure \ref{fig:topog}, $H$ is the wave-free height of the water column including both topographic variations $\DH_2$ and other variations in the mean height, $\DH_1$. Unlike in previous sections, we do not necessarily attribute these variations to geostrophy. 

Providing the fluctuations to mean height $\bar{H}$ are small,
\begin{equation}
    \omf \approx \frac{k_h(gk_h - \omi^2)\DH}{2\omi} + \bU \bcdot \bk.
\end{equation}
For small-amplitude, horizontally isotropic height variations which vary over longer length and time scales than the waves, the azimuthal diffusivity is
\begin{equation}
    \D_{\phi \phi} = \underbrace{\frac{4 \upi k_h^2}{c \vphantom{\omega^2_0}} \int_{0}^{\infty} K_h^2 E(K_h) \, \d K_h}_{\text{Doppler shift}} \, + \, \underbrace{\frac{\upi k_h^2(gk_h - \omi^2)^2}{2 c \omi^2} \int_{0}^{\infty} K_h^2\mathcal{H}(K_h) \, \d K_h}_{\text{height fluctuations}}\label{DijSW3top2},
\end{equation}
which we find by comparing the surface wave dispersion relation \eqref{disp_surf} with that of shallow water waves \eqref{omfSWtop} and switching $g^2 k_h^4 \mathcal{B}(K_h)$ for $k_h^2(gk_h - \omi^2)^2\mathcal{H}(K_h)$ in the topography-induced diffusivity of the rotating shallow water system, \eqref{DijSW3top}. Here, $\mathcal{H}$ is the spectrum of fluctuations in total water depth.
The diffusivity \eqref{DijSW3top2} is valid for a background velocity $\bU$ not correlated with topography or other height-fluctuation effects.


The expressions given in this section are for diffusivities induced by an approximately time-independent $\omf$ which is clearly valid for bottom topography. \citet{Tolman90} considers surface waves propagating through temporally and spatially varying tidal currents which induce variations in depth. We refer the interested reader to \citetalias{cox_kafiabad_vanneste_2023} for work on the time-dependent case.

It should be possible to show the topography-induced diffusivity in \eqref{DijSW3top2} is a limiting case of kinetic equations for Bragg scattering of surface gravity waves by bathymetry such as those derived by \citet{Hasselmann66} and \citet{ARDHUIN_HERBERS_2002}. We leave this proof to future work.






\section{Discussion}\label{sec:discussion}

Scattering by turbulent flow leads to the diffusion of inertia-gravity-wave energy in spectral space. In previous work on the topic  \citep{mccomas_bretherton_1977, kafiabad_savva_vanneste_2019, Savva2020, dong_buhler_smith_2020, cox_kafiabad_vanneste_2023}
the mechanism for spectral wave diffusion is a Doppler shift term in the waves' dispersion relation. In this paper, we argue at the level of the dispersion relation that other mechanisms can cause significant wave diffusion. We provide two \textcolor{black}{main} examples: a fluctuation in the mean height of a rotating shallow water system due to the background geostrophic flow, and an effective variation in buoyancy frequency for the 3D Boussinesq system, caused by vertical background flow buoyancy gradients. There is precedent for this -- \citet{chavanne2010surface} find refraction by vertical buoyancy gradients to be more significant than the Doppler shift effect in the context of internal tides.

We generalise the derivation of \citet{kafiabad_savva_vanneste_2019} to give a diffusion equation valid for any slowly evolving, weak inhomogeneities the waves encounter. For our two examples, we evaluate the corresponding diffusivity.

In the Boussinesq case, we confirm the buoyancy-fluctuation effect can be significant although as waves diffuse to higher wavenumbers, we find this effect is greatly reduced. We also find a reduction for higher frequency waves and background flows with large aspect ratios between horizontal and vertical motions, as is the case in the ocean.

We solve the steady-state diffusion equation and find the resulting energy spectrum agrees with this conclusion -- as the waves move further into the WKB regime, the ratio between buoyancy fluctuation and Doppler shift effects decays to zero.

Scaling predictions from the Boussinesq dispersion relation of the relative importance of Doppler shift and vertical buoyancy gradients differ, with respect to wave frequency, to those found from the complete computation of the diffusivities. This is because large parts of the flow  energy spectrum which do not meet the relevant resonance condition cannot contribute to scattering the waves.

In the shallow water system, we find that for small Burger numbers the height-fluctuation diffusivity is comparable to or larger than the Doppler-shift diffusivity. This is supported by ray simulations which (a) validate the diffusion approximation of the action conservation equation and; (b) confirm the relative magnitude of the height fluctuation and Doppler shift effects. 
The Burger number does not have to be vanishingly small for the height fluctuation effect to be significant -- at $Bu = O(1)$, the two effects differ only by a factor of $1/4$.



Remarkably, the Doppler shift effect is decoupled from the other diffusion mechanisms in the shallow water and Boussinesq diffusion equations. This occurs despite the Doppler shift being related to these other effects through the geostrophic and thermal wind balances. Uncorrelated effects will also be uncoupled from the Doppler shift. For example, shallow water waves \textcolor{black}{ and surface gravity waves scattered by topography, for which the corresponding diffusivities are evaluated in \S\ref{sec:topography}.}


Our work is relevant to internal tides which, alongside near-inertial waves, dominate the IGW energy spectrum \citep*{FerrWun2009}. Previous work uses ray tracing to model internal tide energy distributions and finds good agreement with observation despite marginal scale separation between the tides and background eddies \citep{park2006internal,rainville2006propagation,chavanne2010surface}. Furthermore, if internal tides propagate through a barotropic background flow, they form a set of uncoupled shallow water equations \citep[e.g.][]{savv-vann18}. Beyond this, it is clear that our approach applies for any waves propagating through large-scale inhomogeneities.

A possible application of the formula we obtain for the spectral diffusivity is to the representation of the impact of unresolved turbulence on inertia-gravity waves parameterised in atmosphere and ocean models. In ray-tracing modules such as MS-GWaM \citep{1DMSG-WaMI, 1DMSG-WaMII}, the diffusion could be included by means of additional white-noise terms in the wavevector ray equation, leading to a stochastic parameterisation grounded in the physics of wave--flow interactions.

\textcolor{black}{The diffusion regime is characterised by the weak-interaction assumption of statistical scattering and the WKB approximation. For diffusion induced by Doppler shift, the weak interaction assumption is equivalent to a weak-flow approximation: the group speed of the waves is much greater than the characteristic background flow speed, $c \gg U_*$. Depending on the wave type, this may not hold consistently throughout the evolution. Conditions which ensure that $c \gg U_*$ for IGWs in the Boussinesq system are discussed in appendix A of \citecol{\citet{cox_kafiabad_vanneste_2023}}. \citecol{\citet*{dong2023}} demonstrate the breakdown of the weak-flow assumption in this context.}

\backsection[Acknowledgements]{\textcolor{black}{We thank the three anonymous referees for their valuable comments.} This work was in part undertaken during a research visit by M.R.C to C. Eden and M. Chouksey at the Institut f{\"u}r Meereskunde, Universit{\"a}t Hamburg. We thank both for their hospitality.}

\backsection[Funding]
{M.R.C. is supported by the MAC-MIGS Centre for Doctoral Training under grant EP/S023291/1
of the UK Engineering \& Physical Sciences Research Council (EPSRC). H.A.K. is supported by EPSRC grants 
EP/Y021479/1 and EP/Y032624/1. J.V. is supported by the UK Natural Environment Research Council
grant NE/W002876/1.}

\backsection[Declaration of interests]{The authors report no conflict of interest.}

\backsection[Data availability statement]{\textcolor{black}{The ray tracing code and data that support the findings of this study are available at https://github.com/michaelrcox/inhomogeneity-induced-wavenumber-diffusion.}}


\backsection[Author ORCIDs]{M. R. Cox, https://orcid.org/0000-0002-9329-3644, H. A. Kafiabad, https://orcid.org/
0000-0002-8791-9217; J. Vanneste, https://orcid.org/0000-0002-0319-589X}


\appendix

\section{Action conservation for a shallow water system of variable height}\label{app:wkb}

The conservation equation for wave action density \eqref{actioncons} can be derived for the shallow water model because it is an example of a non-canonical Hamiltonian system, as discussed in \citet{VanShep1999}. 

We apply the WKB analysis of, for example, \citet{Achatz2022} in the context of atmospheric IGWs, to a rotating shallow water system. The shallow water equations for a horizontal fluid layer with velocity $\bu$ and height $h$ are the momentum equation,
\begin{equation}\label{momconswkb}
    \pa{t} \bu + \bu \bcdot \gx \bu + f \be_{z} \times \bu = - g \gx h,
\end{equation}
and the conservation of mass,
\begin{equation}\label{massconswkb}
    \pa{t} h + \gx \bcdot (h \bu) = 0.
\end{equation}
Inline with the notation in figure \ref{fig:SWpaper}, we separate the fluid into a background flow component and a wave component,
\begin{equation}\label{splitting}
    \begin{pmatrix}
        \bu \\
        h
    \end{pmatrix}
    \rightarrow
    \begin{pmatrix}
        \bU\\
        H
    \end{pmatrix}
    +
    \begin{pmatrix}
        \bu \\
        h
    \end{pmatrix},
\end{equation}
where lowercase symbols now indicate wave variables and uppercase the background flow. We apply a WKB ansatz to the wave part,
\begin{equation}\label{wkbansatz1}
    \begin{pmatrix}
        \bu\\
        h
    \end{pmatrix}
    =
    \begin{pmatrix}
        \bu'(\bX, T)\\
        h'(\bX, T)
    \end{pmatrix}
    \e^{\i \Theta(\bX, T)/\epsilon}.
\end{equation}
Here, we introduce the slow time and space scales $(\bX, T) = \epsilon (\bx, t)$, $\epsilon \ll 1$. The background flow $(\bU(\bX, T), H(\bX, T))^{\T}$ and wave amplitudes, hereafter denoted by
\begin{equation}\label{phidef}
    \bphi = (\bu', h')^{\T},
\end{equation}
evolve on these slow scales, whereas the wave phase $\Theta/\epsilon$ evolves $O(1/\epsilon)$ quicker. (Note $\epsilon$ is the ratio between the wavelength and characteristic flow length scale and does not necessarily coincide with the $\epsilon$ of \S\ref{sec:GD}. Likewise, the slow time and space scales here are not the same as those defined in \S \ref{sec:GD} because they are linear, not quadratic, in the small parameter. Upright $\Theta$ is used to distinguish the wave phase from the background flow polar angle, $\iTheta$.) We introduce the local wavevector and frequency
\begin{equation}\label{thetaxt}
    \bk = (k_1, k_2)^\T = \gX \Theta \quad \text{and} \quad \omega = - \pa{T} \Theta,\addtocounter{equation}{1}\tag{\theequation a,b}
\end{equation}
and expand the wave amplitudes in $\epsilon$,
\begin{equation}\label{amplitudeexpansion}
    \bphi(\bX, T) = \bphi_0 + \epsilon \bphi_1 + O(\epsilon^2)
    = \begin{pmatrix}
        \bu_0'\\
        h_0'
    \end{pmatrix}
    + \epsilon
    \begin{pmatrix}
        \bu_1'\\
        h_1'
    \end{pmatrix}
    + O(\epsilon^2).
\end{equation}
In what follows, we drop the primes on the amplitudes $\bu_0', h_0'$ etc. We look at terms linear in the wave phase. 

At $O(1)$ in $\epsilon$, momentum and mass conservation \eqref{momconswkb}--\eqref{massconswkb} become the right and left eigenvector problem
\begin{equation}\label{evecprob}
    \i \mB \mmu \bphi_0 = \i \omin \bphi_0 \quad \text{and} \quad \bphi^\dagger_0 \mmu \mB \mmu = \bphi_0^\dagger \mmu \omin,\addtocounter{equation}{1}\tag{\theequation a,b}
\end{equation}
where
\begin{equation}\label{bmu_matrices}
    \mB = \begin{pmatrix}
        0 & \i f/H & k_1\\
        -\i f/H & 0 & k_2\\
        k_1& k_2 & 0
    \end{pmatrix}
    \quad \text{and} \quad
    \mmu = \begin{pmatrix}
        H & 0 & 0\\
        0 & H & 0\\
        0 & 0 & g
    \end{pmatrix}.\addtocounter{equation}{1}\tag{\theequation a,b}
\end{equation}
The form of the left eigenvector in \eqref{evecprob} is because $\mB$ and $\mmu$ are Hermitian. Here, the intrinsic frequency $\omin$ is the frequency of the waves moving with the background flow,
\begin{equation}\label{intrinsic}
    \omin = \omega - \bU \bcdot \bk.
\end{equation} The eigenvalues are 
\begin{equation}\label{omins}
    \omin = 0, \pm (f^2 + gH k_h^2)^{1/2}.
\end{equation}
The zero eigenvalue corresponds to the background flow mode and the non-zero eigenvalues are waves propagating with the same frequency in opposite directions.

We consider the effect the inhomogeneities have on wave amplitude and thus energy distribution. We choose the positive wave eigenvalue without loss of generality which has the corresponding eigenvector
\begin{equation}\label{evec0}
    \bphi_0 = \frac{\mathscr{a}}{(Hk_h^2)^{1/2}\omin} \begin{pmatrix}
        \i f k_2 + k_1 \omin\\
        -\i f k_1 + k_2 \omin\\
        Hk_h^2
    \end{pmatrix},
\end{equation}
where $\mathscr{a}$ is an $(\bX, \bk, T)$-dependent complex amplitude parameterising the eigenspace of $\mB \mmu$.
The $\mathscr{a}$-independent part of the eigenvector is non-unique in that it can be rotated by a complex phase $\e^{\i \theta}$. The eigenvector \eqref{evec0} is normalised such that the square amplitude $|\mathscr{a}|^2$ corresponds to the energy density, $E(\bX, \bk, T)$,
\begin{equation}\label{normy}
    \frac{1}{2}\bphi_0^\dagger \mmu \bphi_0 =  \underbrace{\frac{H|\bu_0|^2 + g|h_0|^2}{2}}_{E} = |\mathscr{a}|^2.
\end{equation}
We seek an evolution equation for the energy density $E$. 

At $O(\epsilon)$, the momentum and mass conservation equations \eqref{momconswkb}--\eqref{massconswkb} become
\begin{equation}
    \Dv{\bU} \bphi_0 + \begin{pmatrix}
        0 & 0 & \pa{X_1}\\
        0 & 0 & \pa{X_2}\\
        \pa{X_1} & \pa{X_2} & 0
    \end{pmatrix}
    \mmu \bphi_0 +\begin{pmatrix}
        \pa{X_1} U & \pa{X_2} U & 0\\
        \pa{X_1} V & \pa{X_2} V & 0\\
        0 & 0 & \gX \bcdot \bU
    \end{pmatrix}
    \bphi_0 + \i(\mB \mmu - \omin)\bphi_1 = 0,
\end{equation}
where 
\begin{equation}\label{advecderiv}
    \Dv{\bU} = \pa{T} + \bU \bcdot \gX,
\end{equation}
is the advective derivative with $\bU$. We pre-multiply by the left eigenvector $\bphi_0^\dagger \mmu$ to remove the $\bphi_1$ term leaving
\begin{equation}\label{bigeqn}
    \underbrace{\vphantom{\begin{pmatrix}
        \pa{x} U & \pa{y} U & 0\\
        \pa{x} V & \pa{y} V & 0\\
        0 & 0 & \gx \bcdot \bU
    \end{pmatrix}}\bphi_0^\dagger \mmu\Dv{\bU}\bphi_0}_{\mcirc{1} } + \underbrace{\vphantom{\begin{pmatrix}
        \pa{X_1} U & \pa{X_2} U & 0\\
        \pa{X_1} V & \pa{X_2} V & 0\\
        0 & 0 & \gx \bcdot \bU
    \end{pmatrix}}\bphi_0^\dagger \mmu\begin{pmatrix}
        0 & 0 & \pa{X_1}\\
        0 & 0 & \pa{X_2}\\
        \pa{X_1} & \pa{X_2} & 0
    \end{pmatrix}
    \mmu \bphi_0}_{\mcirc{2}} +\underbrace{\bphi_0^\dagger \mmu\begin{pmatrix}
        \pa{X_1} U & \pa{X_2} U & 0\\
        \pa{X_1} V & \pa{X_2} V & 0\\
        0 & 0 & \gX \bcdot \bU
    \end{pmatrix}\bphi_0}_{\mcirc{3}} = 0.
\end{equation}
We add the complex conjugate, divide by 2 and evaluate each term. Expanding term 1 and rearranging gives
\begin{equation}
    \frac{1}{2}(\mcirc{1} + \cc) = \frac{1}{2} \left(\Dv{\bU}(H|\bu_0|^2 + g|h_0^2|) - |\bu_0|^2 \Dv{\bU}H\right) = \frac{1}{2} \Dv{\bU}E - \frac{1}{2}|\bu_0|^2 \Dv{\bU}H,
\end{equation}
where we use the normalisation \eqref{normy} for the second equality. Using the product rule, term 2 becomes
\begin{equation}
    \frac{1}{2}(\mcirc{2} + \cc)
    = \gX \bcdot (gH (\bu^*_0 h_0 + \bu_0 h_0^*)/2) = \gX \bcdot (\bcin E),
\end{equation}
where we use the explicit form of the eigenvector \eqref{evec0} to find
\begin{equation}
    \frac{1}{2}(\bu^*_0 h_0 + \bu_0 h_0^*) = \bk E/\omin = \bcin E/(gH).
\end{equation}
Here, we have introduced
\begin{equation}\label{groupvel}
    \bcin = \left( \gk \omin \right)_{\bX, T} = \frac{gH \bk}{\omin},
\end{equation}
the group velocity associated with the waves' intrinsic frequency. This differs from $\bc$ \eqref{lo-gc}, the group velocity associated with the waves' bare frequency $\omi$. The $\bX, T$ - subscript indicates that these variables are kept constant whilst taking the partial derivative, despite both being implicitly dependent on $\bk$.

The final term is
\begin{align}\notag
    \frac{1}{2}\left(\mcirc{3} + \cc\right)
    &= (H |\bu_0|^2 + g |h_0|^2) \gX \bcdot \bU - H (|u_0|^2 \pa{X_2}V + |v_0|^2 \pa{X_1}U)\\&\hspace{5cm}+\frac{1}{2}H(v_0u_0^* + v_0^* u_0)(\pa{X_1}V + \pa{X_2}U)\\
    &= E \left(2\gX \bcdot \bU - \frac{\omin^2 - g H k_2^2}{\omin^2} \pa{X_2} V - \frac{\omin^2 - g H k_1^2}{\omin^2} \pa{X_1} U + \frac{gHk_1k_2}{\omin^2}(\pa{X_1}V + \pa{X_2}U)\right)\label{usquare}\\
    &= E\left(\omin \gX \bcdot \bU + \bk \bcdot \left(\bcin \bcdot \gX \bU \right)\right)/\omin,
\end{align}
where we use the normalisation \eqref{normy} and
\begin{equation}\label{u2v2etc}
    |u_0^2| = E\frac{\omin^2 - gHk_2^2}{H\omin^2}, \quad |v_0^2| = E\frac{\omin^2 - gHk_1^2}{H\omin^2}, \quad \text{and} \quad v_0 u_0^* + v_0^* u_0^* = E\frac{2 g k_1 k_2}{\omin^2},\addtocounter{equation}{1}\tag{\theequation a,b,c}
\end{equation}
derived from the eigenvector \eqref{evec0}. All together, \eqref{bigeqn} becomes
\begin{align}
    0 &= \Dv{\bU}(E) - \frac{1}{2}|\bu_0|^2 \Dv{\bU} H + \gX \bcdot (\bcin E) + E\left(\omin \gX \bcdot \bU + \bk \bcdot \left(\bcin \bcdot \gX \bU \right)\right)/\omin\\
    &= \omin\left(\pa{T}A + \gX \bcdot (\bc_g A)\right) + A\left( \pa{T} + \bc_g \bcdot \gX\right)\omin - \frac{1}{2}|\bu_0|^2 \Dv{\bU} H + A\bk \bcdot \left(\bcin \bcdot \gX \bU \right),\label{thirty}
\end{align}
where
\begin{equation}\label{totgroupvel}
    \bc_g = \bcin + \bU,
\end{equation}
is the total group velocity and
\begin{equation}\label{actionamp}
    A = E/\omin,
\end{equation}
is wave action -- the energy density normalised by the intrinsic frequency.

We introduce
\begin{equation}\label{totfreqexplicit}
\textcolor{black}{\Omega}(\bX, \bk(\bX, T), T) = \omega(\bX, T),
\end{equation}
the total frequency with explicit $\bk$ dependence.
The eikonal equations are derived from cross-derivatives of \eqref{thetaxt},
\begin{equation}\label{eikonal1}
    (\pa{T} + \bc_g \bcdot \gX) \omega \equiv \pa{T} \textcolor{black}{\Omega}
    = \bk \bcdot \pa{T} U + (\pa{H}\omin )_{\bk} \pa{T} H,
\end{equation}
and
\begin{align}\label{eikonal2}
    (\pa{T} + \bc_g \bcdot \gX) \bk \equiv - \gX \textcolor{black}{\Omega}
    = -\bk \bcdot (\gX \bU) - (\pa{H}\omin )_{\bk}\gX H.
\end{align}
To find derivatives of $\textcolor{black}{\Omega}$, we have used the total frequency $\omega$, defined through \eqref{intrinsic}--\eqref{omins}. Note that $(\pa{H} \omin)_{\bk}$ is the partial derivative of $\omin$ with respect to $H$ with $\bk$ fixed. Then, using \eqref{eikonal1}--\eqref{eikonal2},
\begin{align}
    \left( \pa{T} + \bc_g \bcdot \gX\right)\omin &= \left( \pa{T} + \bc_g \bcdot \gX\right)\omega - \bU \bcdot \left( \pa{T} + \bc_g \bcdot \gX\right)\bk - \bk \bcdot \left( \pa{T} + \bc_g \bcdot \gX\right) \bU\\
    &= (\pa{H}\omin)_{\bk} \pa{T}H + (\pa{H}\omin)_{\bk}\bU\bcdot  \gX H + \bk \bcdot (\bU \bcdot \gX \bU) - \bk \bcdot (\bc_g \bcdot \gX \bU)\label{A33}\\
    &= (\pa{H} \omin)_{\bk} \Dv{\bU} H-  \bk \bcdot (\bcin \bcdot \gX \bU).
\end{align}
Substituting this result into \eqref{thirty} yields
\begin{align}
    \omin\left(\pa{T}A + \gX \bcdot (\bc_g A)\right) + A(\pa{H} \omin)_{\bk} \Dv{\bU} H - \frac{1}{2}|\bu_0|^2 \Dv{\bU} H = 0.
\end{align}
If the background flow is non-divergent flow, as is the case for the geostrophic flow we consider, then $\Dv{\bU} H = 0$ and, letting our book-keeping parameter $\epsilon = 1$,
\begin{equation}\label{actionnofterm}
    \pa{t}A + \gx \bcdot (\bc_g A) = 0.
\end{equation}
Otherwise, we have that
\begin{equation}\label{actionfterm}
    \pa{t}A+ \gx \bcdot (\bc_g A) - \frac{f^2}{\omin^2} A \gx \bcdot \bU = 0,
\end{equation}
where we use the square components \eqref{u2v2etc} and
\begin{equation}
    \pa{t} H + \bU \bcdot \gx H = - H \gx \bcdot \bU,
\end{equation}
which holds because the flow is a solution of the rotating shallow water equations \eqref{momconswkb}--\eqref{massconswkb}.

For rotation effects to be significant in the dispersion relation of the shallow water waves, $1 \sim Bu (k_h/K_*)^2$ where $Bu$ is the flow Burger number as defined by \eqref{burgereqn}. Then, $f^2/\omin^2 = O(1)$. By the WKB ansatz, $k_h/K_* \gg 1$ and so $Bu \ll 1$. This is the planetary-geostrophy regime whereby the scale of wave motion is much greater than the Rossby radius of deformation and the background flow is in geostrophic balance \citep[see, for example,][]{Vallis_2017}. Thus, the flow is non-divergent and \eqref{actionfterm} reduces to \eqref{actionnofterm}. If rotation effects are neglected, we have gravity waves and $f^2/\omin^2 \ll 1$. Therefore, in the WKB regime, the divergence term in \eqref{actionfterm} can always be neglected. 

A consistent derivation for waves propagating with significant rotation $f$ in a QG flow $Bu \sim O(1)$ is not possible. We defer to \citet*{bretherton1971general}: ``\dots when the physical situation is inappropriate, no amount of juggling will give a consistent, slowly varying wavetrain."

The conservation law \eqref{actionnofterm} is for wave action $A(\bx, t)$ defined by \eqref{actionamp}. It is coupled to the eikonal equation for $\bk$, \eqref{eikonal2}. For a conservation law that spans all of $(\bx,\bk)$ space, we index each solution of \eqref{eikonal2} and \eqref{actionnofterm} with the \textcolor{black}{2D} parameter $\balpha$ such that
\begin{equation}\label{labelledsols}
    \pa{t}A_{\balpha} + \gx \bcdot (\bc_{g\balpha} A_{\balpha}) = 0 \quad \text{and} \quad (\pa{t} + \bc_{g\balpha} \bcdot \gx) \bk_{\balpha} = - \gx \textcolor{black}{\Omega}_{\balpha},\addtocounter{equation}{1}\tag{\theequation a,b}
\end{equation}
and introduce the wave action density,
\begin{equation}\label{actionintegral}
    a(\bx, \bk, t) = \int A_{\balpha}(\bx, t)\delta(\bk - \bk_{\balpha}) \d \balpha,
\end{equation}
a superposition of the individual solutions to the coupled equations. By taking the time derivative of \eqref{actionintegral} and applying the coupled equations $\eqref{labelledsols}$, the conservation equation for wave action \eqref{actioncons} with total frequency \eqref{SWtot_2} is obtained. The steps between \eqref{labelledsols}--\eqref{actionintegral} and \eqref{actioncons} are standard and do not rely on the specific form of the dispersion relation and are omitted here for brevity. See, for example, \S 10.3.8 of \citet{Achatz2022}. We emphasise that the height $H(\bx, t)$ is taken to vary on the same physical and temporal scales as the flow throughout this derivation.

\section{Boussinesq diffusivities}
\subsection{Obtaining diffusivity expressions}\label{app:boussdiff}

In this appendix, we evaluate the general diffusivity \eqref{general_diff3} for the Boussinesq system with vertical buoyancy gradients. 

We first substitute $\omf$, defined in \eqref{omtotBous2} with $B = f \pa{z} \psi$ by the thermal wind balance \eqref{thermalwind}, into the correlation function \eqref{lambda}. Due to the two-dimensional nature of geostrophic flow, the Doppler shift term is the same here as it is in \eqref{lambdaSW2}, the shallow water case, with the skew gradient 2D as before. The cross terms also cancel by a similar argument to \eqref{crossterms}. Then,
\begin{equation}
    \Lambda(\by,\bk, r)= \underbrace{\vphantom{\frac{f^2 \sin^4 \theta}{4 \omi^2}}-(\bk \bcdot \gx^{\perp})^2\langle  \psi(\bx, t) \psi(\bx + \by, t + r)\rangle}_{\text{Doppler shift}}
    \, + \, \underbrace{\frac{f^2 \sin^4 \theta}{4 \omi^2} \pa{zzzz}\langle \psi(\bx, t)\psi(\bx + \by, t + r)\rangle}_{\text{buoyancy fluctuation}}.
\end{equation}
The vertical derivative in the buoyancy fluctuation term has been moved outside the ensemble average using integration by parts and the symmetry of $\bx$ and $\by$ arguments. In Fourier space, this becomes
\begin{equation}
    \hat{\Lambda}(\bK,\bk, \iOmega) = |\bk_h \times \bK_h|^2 E_\psi(\bK, \iOmega) \, + \, \frac{f^2 \sin^4 \theta K_v^4}{4 \omi^2} E_\psi(\bK, \iOmega).
\end{equation}
 Here, $K_v$ is the flow's vertical wavenumber and $\bK_h$ its horizontal wavevector (distinct from $K_h = |\bK_h|$, the horizontal wavenumber). Switching to polar coordinates $\bK = (K, \gamma, \iTheta)$, $\gamma$ as before the angle between the horizontal wavevectors $\bK_h$ and $\bk_h$,
 \begin{equation}\label{lambdahatBouss}
    \hat{\Lambda}(\bK,\bk, \iOmega) = 2 k^2 \sin^2 \theta \sin^2 \gamma E(\bK, \iOmega) \, + \, \frac{f^2 \sin^4 \theta K^2 \cos^4 \iTheta}{2 \omi^2 \sin^2 \iTheta} E(\bK, \iOmega).
\end{equation}
We have used the same definitions of $E_\psi$ and $E$ as in the shallow water case, but note that $E = K_h^2 E_\psi/2 = K^2 \sin^2 \iTheta E_\psi/2$. Combined with \eqref{general_diff3}, we have a diffusivity that accounts for buoyancy gradients in a time-dependent flow.


We simplify to a time-independent flow and substitute \eqref{lambdahatBouss} into \eqref{timeindFTdiff},
\begin{align}
    \D_{ij} = &2\upi k^2 \sin^2 \theta \int_{0}^{\infty} \d K \int_0^{\upi}\d\iTheta \int_{-\upi}^{\upi}\d\gamma \, K_i K_j K^2 \sin\iTheta \sin^2 \gamma E(\bK)\delta(\bK \bcdot \bc)\notag\\
    &+ \frac{\upi f^2 \sin^4 \theta }{2 \omi^2 }  \int_{0}^{\infty} \d K \int_0^{\upi}\d\iTheta \int_{-\upi}^{\upi}\d\gamma \, \frac{K_i K_j K^4 \cos^4 \iTheta E(\bK) \delta(\bK \bcdot \bc)}{\sin \iTheta}.
\end{align}
As in \citetalias{kafiabad_savva_vanneste_2019} and \citetalias{cox_kafiabad_vanneste_2023}, we expand $\boldsymbol{K}$ in the local spherical basis $(\boldsymbol{e}_k, \boldsymbol{e}_\theta, \boldsymbol{e}_\phi)$ associated with $\boldsymbol{k}$ such that
\begin{equation}\label{K}
    \boldsymbol{K} = K \sin \mathit{\Theta}\left( (\sin \theta \cos \gamma + \cot \mathit{\Theta} \cos \theta) \boldsymbol{e}_k + (\cos \theta \cos \gamma - \cot \mathit{\Theta} \sin \theta)\boldsymbol{e}_\theta +\sin \gamma \boldsymbol{e}_\phi\right).
\end{equation}
We consider $\bD$ in spherical components i.e.\ $\D_{\theta \theta} = \be_\theta \bcdot \bD \bcdot \be_\theta$, $\D_{\theta k} = \be_\theta \bcdot \bD \bcdot \be_k$ etc. We see that $\bD \bcdot \bc = 0$ because upon moving $\bc$ inside the integral, each integrand contains a factor of $\bK \bcdot \bc$, the argument of the delta function. As $\bK \bcdot \bc = c \bK \bcdot \boldsymbol{e}_\theta$ because $\bc = \gk \omi(\theta)$, this means that $\D_{\theta \theta}$, $\D_{\theta \phi} = \D_{\phi \theta}$ and $\D_{\theta k} = \D_{k \theta}$ are all zero.

As in \citetalias{cox_kafiabad_vanneste_2023}, we use parity arguments to show that $\D_{\phi k} = \D_{k \phi} = 0$. The parity of $K_iK_j$ with respect to $\gamma$ is determined by the parity of pairwise products of $\boldsymbol{e}_k \bcdot \bK$ and $\boldsymbol{e}_\phi \bcdot \bK$. The parity of the delta function is even because of the parity of $\bK \bcdot \be_\theta$. Thus, only $\mathsfi{D}_{kk}$ and $\mathsfi{D}_{\phi \phi}$ have even integrands in $\gamma$ and only these components are non-zero.
We assume the energy spectrum is horizontally isotropic such then $E(\bK) = E(K, \iTheta)$. This enables integration with respect to $\gamma$ using, for example, a substitution of $\xi = \cos \gamma$. Thus,
\begin{align}
    \D_{ij} = &\frac{4\upi k^2 \sin^2 \theta}{c |\cos \theta|} \int_{0}^{\infty} \d K \int_{\theta}^{\upi - \theta}\d\iTheta\, K_i K_j(\xi_*) K (1 - \xi_*^2)^{1/2}E(K, \iTheta)\notag\\
    &+ \frac{\upi f^2 \sin^4 \theta }{ \omi^2 c |\cos \theta|}  \int_{0}^{\infty} \d K \int_{\theta}^{\upi - \theta}\d\iTheta \, \frac{K_i K_j(\xi_*) K^3 \cos^4 \iTheta E(K, \iTheta)}{\sin^2 \iTheta (1 - \xi_*^2)^{1/2}},
\end{align}
where $\xi_* = \cot \mathit{\Theta}/\cot\theta$. Note only values of $\mathit{\Theta}$ for which $|\cot \mathit{\Theta}/\cot \theta| < 1$ contribute to the integral, which reduces the integration range  to $(\theta, \upi - \theta)$. This is discussed in \S\ref{sec:GDB}.

We evaluate $(\be_k \bcdot \bK)^2$ and $(\be_\phi \bcdot \bK)^2$ at $\cos \gamma = \xi_*$,
\begin{equation}
    (\be_k \bcdot \bK)^2|_{\xi_*} = K^2 \sin^2 \iTheta (\sin \theta \xi_* + \cot \mathit{\Theta} \cos \theta)^2\quad \text{and} \quad (\be_\phi \bcdot \bK)^2|_{\xi_*} = K^2 \sin^2 \iTheta (1 - \xi_*^2),
\end{equation}
and use $c = (\pa{\theta} \omega_0) /k = (\bar{N}^2 - f^2) \sin \theta \cos \theta /(k \omega_0)$ \textcolor{black}{ to give the two non-zero components of $\bD$, \eqref{DkkBouss}--\eqref{DphiphiBouss}}.
Reassuringly, the Doppler shift terms in both these components agree with (A13) in \citetalias{kafiabad_savva_vanneste_2019}, up to a $(2 \upi)^3$ factor due to differing Fourier convention, and typographical errors in both the lower limits of the integrals. One of these errors is corrected in (2.11) of \citetalias{cox_kafiabad_vanneste_2023}, but the lower limit of the $K$ integral is still incorrect.

\subsection{Frequency dependence of the radial diffusivity ratio}\label{app:RBpowers}
We explain the $(\omi/f)^{-2}$ power law of ratio $\RBpk$ \eqref{rbprimes} and its behaviour as $\theta \rightarrow 0$. Substituting the radial diffusivity components \eqref{DkkBouss} into \eqref{rbprimes},
\begin{equation}
    \RBpk = \frac{\sin^4\theta}{4(\omi/f)^2 k_h^2} \frac{\int_0^\infty \int_\theta^{\upi-\theta} K^5 \cos^6 \iTheta E(K, \iTheta) (1 - \cot^2 \iTheta /\cot^2 \theta)^{-1/2} \sin^{-2}\iTheta \, \d K \d \iTheta}{\int_0^\infty \int_\theta^{\upi-\theta} K^3 \cos^2 \iTheta E(K, \iTheta) (1 - \cot^2 \iTheta /\cot^2 \theta)^{1/2} \, \d K \d \iTheta}.
\end{equation}
As $\omi/f \rightarrow 1$ and $\theta \rightarrow 0$, the prefactor goes to zero and the integrals tend to spectrum-dependent constants. Thus, the ratio $\RBpk$ tends to zero.

Transforming to variables $(K_h, \textcolor{black}{\xi_*}) = (K/\sin\iTheta, \cot \iTheta/\cot \theta)$, we have that,
\begin{equation}\label{appendixratio}
    \RBpk = \frac{\cos^4 \theta}{4 (\omi/f)^2k_h^2}\frac{\int_{0}^{\infty}  K_h^4 \d K_h\int_{-1}^{1} \textcolor{black}{\xi_*}^6 \tilde{E}(K_h, K_v = \textcolor{black}{\xi_*} K_h \cot \theta) (1 - \textcolor{black}{\xi_*}^2)^{-1/2} \, \d \textcolor{black}{\xi_*}}{\int_{0}^{\infty} K_h^2 \d K_h \int_{-1}^{1}\textcolor{black}{\xi_*}^2 \tilde{E}(K_h, K_v = \textcolor{black}{\xi_*} K_h \cot \theta)(1 - \textcolor{black}{\xi_*}^2)^{1/2} \, \d \textcolor{black}{\xi_*}}.
\end{equation}
Here, we introduce the cylindrical energy spectrum
\begin{equation}\label{cylindrical}
    \textcolor{black}{\tilde{E}(K_h, K_v) = 2 \upi K_h E(\bK).}
\end{equation}
For an energy spectrum which is vertically homogeneous across the integration domain in spectral space, $\tilde{E}(K_h, K_v) \approx \tilde{E}(K_h)$ and both integrals can be evaluated with respect to $\textcolor{black}{\xi_*}$. Then,
\begin{equation}\label{appendixratio2}
    \RBpk = \frac{5 \cos^4 \theta}{8 (\omi/f)^2k_h^2}\frac{\int_{0}^{\infty}  K_h^4 \tilde{E}(K_h) \, \d K_h}{\int_{0}^{\infty}  K_h^2 \tilde{E}(K_h) \, \d K_h} \approx  \frac{5 \cos^4 \theta}{8 (\omi/f)^2(k_h/K_*)^2}.
\end{equation}
For small $\theta$, this gives a $(\omi/f)^{-2}$ power law.

The spectrum can be considered vertically homogeneous, even for small $\theta$, because of the large aspect ratio of the flow, $\alpha$ \eqref{aspect}. This approximation improves as either $\theta$ or $\alpha$ grows, the former because the integration domain over $\iTheta$ shrinks.  For the flow of \citetalias{cox_kafiabad_vanneste_2023}, we find \eqref{appendixratio2} to be a good estimate of $\RBpk$ for $\cot \theta \gtrsim \alpha$, \textcolor{black}{the aspect ratio of the flow \eqref{aspect}}, i.e.\ the point at which the integration domain $(\theta, \upi - \theta)$ coincides with the characteristic wavevector of the flow. At this point, $\theta$ is not large enough for the spectrum to appear homogeneous in the vertical and we attribute this better-than-expected approximation to the prefactor of \eqref{appendixratio} varying more quickly than the ratio of integrals with $\theta$.

\section{\textcolor{black}{Key symbols}}\label{app:glossaries}

This appendix consists of two tables of symbol definitions. Table \ref{tab:symbols} contains key symbols found in the main body of the paper. Table \ref{tab:symbols2}, contains key symbols found only in the preceding appendices. In general, wave variables with flow-associated counterparts are lowercase versions of the flow variables e.g. wavevector $\bk$ and flow wavevector $\bK$ with spherical components $(k, \theta, \phi)$ and $(K, \iTheta, \mathit{\Phi})$ respectively.

\begin{table}
    \centering
    \begin{tabular}{c|c}
        $\omega$ & absolute wave frequency\\
        $\omi$ & frequency in the absence of inhomogeneities, called the \textit{bare} frequency, def.\ \eqref{totfreq}\\
        $\omf$ & inhomogeneity-induced part of $\omega$, def.\ \eqref{totfreq}\\
        $H$ & wave-independent height of the rotating shallow water layer (see figure \ref{fig:SWpaper})\\
        $\DH$ & fluctuations to mean height $\bar{H}$ induced by geostrophy (see figure \ref{fig:SWpaper}(b), def.\ \eqref{geobalSW})\\
        $h/*$ subscripts & denote the horizontal component/characteristic value of a quantity\\
        $Bu, Ro$ & Background flow Burger and Rossby numbers, def.\ \eqref{burgereqn} and \eqref{rossbyeqn}\\
        $L_D$ & Rossby radius of deformation $Bu^{1/2}/K_*$\\
        $L_*, H_*, \alpha$ & background flow characteristic horizontal and vertical length scales, and ratio $L_*/H_*$ def.\ \eqref{aspect}\\
        $\Rsw$ & relative importance of RSW height fluctuation and Doppler shift in $\omf$ (see figure \ref{fig:ratio2}, def.\ \eqref{ratioSW})\\
        $\Rsw'$ & $\Rsw^2$, revised relative importance (see figure \ref{fig:ratio2}, def.\ \eqref{Rswprime})\\
        $\RB$ & relative importance of buoyancy fluctuation to Doppler shift (see figure \ref{fig:ratioBcontour}, def.\ \eqref{ratioB}--\eqref{ratioB2})\\
        $\RB'$ & revised relative importance of buoyancy fluctuation and Doppler shift  def.\ \eqref{rbprimes}\\
        $\bc$ & leading order contribution to group velocity of the waves, def.\ \eqref{lo-gc}, with magnitude $c$\\
        $\bD$ & diffusion tensor with components $\D_{ij}$ as given in \eqref{diffusivity1}\\
        $\Lambda$ & correlation function of $\omf$, def.\ \eqref{lambda}\\
        $\gamma$ & angle between the horizontal components of $\bK$ and $\bk$\\
        $E_\psi$ & stream function power spectrum i.e.\ Fourier transform of $\langle \psi(\bx, t)\psi(\bx + \by, t + r)\rangle$, def.\ \eqref{streamfuncspec}\\
        $E(\bK, \iOmega)$ & background flow kinetic energy spectrum, def.\ \eqref{KineticSpec}\\
        $E(\bK)$ & $E(\bK, \iOmega)$ marginalised over frequencies ($E(K_h)$ or $E(K,\iTheta)$ if horizontally isotropic)\\
        $\mu, \tilde{\mu}$ & directional diffusivity and non-dimensional counterpart, def.\ \eqref{gradient} and \eqref{mutilde}\\
        $k_*$ & forcing wavenumber of waves in \S\ref{sec:f-e-spec}\\
        $Q, P$ & $k$-independent part of Boussinesq Doppler shift and buoyancy fluctuation diffusivities, def.\ \eqref{Qeqn}\\
        $\beta$ & $P/Q$\\
        $\DH_2$ & topagraphic variation in height, see \eqref{DH12}\\
        $\mathcal{B}(\bK)$ & topography spectrum, def.\ \eqref{topspec}\\
        $\mathcal{H}(\bK)$ & spectrum of fluctuations in total water depth,  Fourier transform of $\langle \DH(\bx, t) \DH(\bx + \by, t + r)  \rangle$
        
    \end{tabular}
    \caption{Key symbols in the main text.}
    \label{tab:symbols}
\end{table}

\begin{table}
    \centering
    \begin{tabular}{c|c}
        $\Theta/\epsilon$ & wave phase (see \eqref{wkbansatz1})\\
        $\bphi$ & vector of wave amplitudes $(\bu' ,h')^\T$, def.\ \eqref{phidef}\\
        
        
        $\mB, \mmu$ & Hermitian matrices def. \eqref{bmu_matrices}
        which multiply to give eigenvector problem \eqref{evecprob}\\
        $\omega_{\text{in}}$ & intrinsic frequency, def.\ \eqref{intrinsic}\\
        $\mathscr{a}$ & complex amplitude parameterising eigenspace of $\mB \mmu$, see \eqref{evec0}\\
        $E$ & energy density, def.\ \eqref{normy}\\
        $\Dv{\bU}$ & advective derivative with $\bU$, def.\ \eqref{advecderiv}\\
        $\bcin$ & group velocity associated with the waves' intrinsic frequency, def.\ \eqref{groupvel}\\
        $\left( \gk \omin \right)_{\bX, T}$ & subscript indicates $\bX,T$ are kept constant despite both being implicitly dependent on $\bk$\\
        $\bc_g$ & total group velocity, def.\ \eqref{totgroupvel}\\
        $A$ & wave action, def.\ \eqref{actionamp}\\
        $\Omega$ & total frequency with explicit $\bk$ dependence, def.\ \eqref{totfreqexplicit}\\
        $\balpha$ & 2D parameter indexing solutions of \eqref{eikonal1} and \eqref{eikonal2}\\
        $\xi$ & $\cos \gamma$\\
        $\xi_*$ & $\cos \gamma = \cot \iTheta/\cot \theta$\\
        $\tilde{E}$ & cylindrical background flow energy spectrum, def. \eqref{cylindrical}

    \end{tabular}
    \caption{Key symbols appearing only in the appendices.}
    \label{tab:symbols2}
\end{table}

\bibliographystyle{jfm}
\bibliography{jfm}


\end{document}